\documentclass[11pt]{article}

\usepackage[T1]{fontenc}
\usepackage[utf8]{inputenc}
\usepackage{lmodern}
\usepackage[a4paper,left=16.5mm,right=16.5mm,top=20mm,bottom=22mm]{geometry}
\usepackage{amsmath,amssymb,mathtools,bm}
\usepackage{graphicx}
\usepackage{booktabs,tabularx}
\usepackage[round,authoryear]{natbib}
\usepackage{microtype}
\usepackage[hidelinks]{hyperref}

\hypersetup{
  pdftitle={An Interpretable Low-Rank State-Space Model for Multi-Horizon Simulation of Large-Scale Regional Temperature Fields},
  pdfauthor={Varun Kotharkar and Michael L. Stein}
}
\title{An Interpretable Low-Rank State-Space Model for Multi-Horizon Simulation of Large-Scale Regional Temperature Fields}
\author{Varun Kotharkar\thanks{Corresponding author: \texttt{vsk34@stat.rutgers.edu}} \quad Michael L. Stein\\[0.5em]
\small Department of Statistics, Rutgers University, Piscataway, NJ, USA}
\date{July 2026}

\begin{document}
\maketitle
\begin{abstract}\label{sec:abs}
We propose an interpretable low-rank state-space model for conditional
simulation of daily regional temperature fields. The central statistical
question is whether empirical orthogonal functions (EOFs) can be treated as
stable large-scale features of the field rather than only as sample-dependent
basis vectors for dimension reduction. The framework represents the dominant
temperature field through a small number of retained EOF coefficients, models
their seasonal mean and variance structure, propagates the resulting
low-dimensional state with stable multivariate dynamics, and uses structured
innovations to represent the remaining uncertainty. The resulting reduced-rank
model is interpretable, computationally efficient, and suitable for iterative
ensemble generation. It is designed to support both inference on the structure
of the retained temperature state and prediction through multi-horizon
probabilistic field simulation.

We evaluate the method using daily 2\,m ERA5 temperature over five climatically
distinct U.S.\ regions, each represented on a fixed $64\times64$ grid, under a
strict rolling-origin design. The inferential target is the projected
predictive law of the retained rank-$K^*$ field rather than the full
native-rank temperature field. Across regions, the retained $K^*=4$ EOF state
explains about 96.7--97.5\% of centered-field variance, remains stable under
rolling refits, and identifies coherent large-scale modes with clear
climatological interpretation. The one-step simulations show large
short-lead gains from conditioning on the observed origin state. At 3--14 days,
the dynamical models continue to improve multivariate field scores relative to
climatology. At 30--60 and 90--120 days, trajectory-level hybridization toward
state-matched climatological continuations improves energy and variogram scores
for all three dynamical bases. The results identify a horizon-dependent
trade-off rather than a single universally dominant simulator: hybrid models
provide the strongest field-level scores at medium and long leads, while
climatological resampling remains a strong marginal warm-tail benchmark. The
paper shows how reduced-rank state-space modeling can link dimension reduction,
interpretation, diagnostics, and probabilistic simulation for high-dimensional
environmental fields.
\end{abstract}


\section{Introduction}
\label{sec:intro}

Many applications in climate services, infrastructure planning, energy systems,
agriculture, and risk analysis require ensembles of daily temperature fields
that are credible jointly across space and time, rather than point forecasts at
a few locations. In energy and grid planning, temperature-driven demand and
weather-dependent supply variability require distributional
inputs to quantify peak risk and resource adequacy
\citep{staffell2023demand}. In heat-risk and public-health analysis, health
outcomes respond nonlinearly and with lags to temperature exposure, so that
ensembles preserving persistence, spatial coherence, and upper-tail behavior
can support burden estimation \citep{gasparrini2015mortality}; the spatial
structure of heat waves and their compound interaction with droughts has strong
socioeconomic impact \citep{szemkus2024heatdrought}. In agriculture and food
security, weather generators remain central tools for yield-risk analysis
because changes in variability can matter more than changes in the mean for
crop impacts \citep{semenov1997wg}. In hydrology and water-resources planning,
coherent space-time forcing fields are critical because hydrologic models are
sensitive to persistence and spatial covariance
\citep{clark2004schaake,nguyen2024nsrwg,steinschneider2019vulnerability}. In
insurance and financial risk, calibrated probabilistic hazard characterization
and large ensembles are needed for spatial aggregation and portfolio-level risk
summaries \citep{jewson2025extremerisk}. Across all of these domains, the statistical
task is not simply to predict a single future value, but to construct a
probabilistic representation of the evolving regional temperature field that
preserves its dominant spatial structure, seasonal modulation, and temporal
dependence over multiple lead times
\citep{wilks2011statmethods,gneiting2007scoring,scheuerer2015variogram}.

Stochastic weather generators were developed precisely to reproduce the
long-run statistical behavior of daily weather sequences and remain central to
climate-impact assessment \citep{richardson1981weather,semenov1997wg,
ailliot2015wg_overview}. Extending such methods from single sites to large
regional fields is substantially more difficult. Annual seasonality induces
strong nonstationarity in both marginal distributions and dependence
structure; gridded temperature fields are high dimensional; and even modest
errors in the innovation model can accumulate under iterative simulation,
producing unrealistic dispersion, persistence, or spatial dependence at medium
and long lead times \citep{wilks1998multisite,clark2004schaake,
schefzik2013ecc,brunner2021multisite}.

A central empirical feature of regional temperature fields is that much of
their variability is spatially smooth and highly correlated. This suggests a
reduced-rank representation in empirical orthogonal function (EOF) space:
rather than modeling the full gridded field directly, one models a
lower-dimensional state formed by the dominant coefficient amplitudes
\citep{jolliffe2002pca,hannachi2007eofreview}. In the present setting,
however, reduced rank is not merely a computational device. It is also an
inferential choice. If the retained EOF subspace is stable under repeated
rolling refits, then the leading coefficients may be interpreted as amplitudes
of persistent, physically meaningful regional modes rather than as arbitrary
basis coordinates. Subspace stability, eigenvalue separation, and sampling
variability therefore become central statistical questions rather than
secondary diagnostics \citep{north1982sampling,davis1970rotation,
stewart1990matrix}.

After projection into coefficient space, conditional simulation becomes a
multivariate time-series problem. A natural starting point is a vector
autoregression, which can capture serial dependence within retained modes and
cross-lag dependence across modes \citep{hamilton1994ts,lutkepohl2005var}.
Because the model is refit repeatedly in moderate dimension, however,
unrestricted dynamics can be unstable and estimation noise can materially
affect multi-step propagation. Regularization is therefore essential, both for
statistical robustness and for well-behaved iterative simulation
\citep{doan1984forecast,litterman1986bvar,banbura2010largebvar}.

A second challenge is that an adequate conditional mean model does not by
itself guarantee realistic ensembles. Residual conditional heteroskedasticity
and contemporaneous cross-mode dependence are common in multivariate time
series, and homoskedastic Gaussian innovations may be too restrictive for
multi-step simulation \citep{engle1982arch,bollerslev1986garch,engle2002dcc}.
In our framework, the innovation model has a deliberately limited but important
role: it is introduced only after deterministic seasonal structure and linear
state dynamics have been removed, and is used to represent the residual
uncertainty that remains.

A third issue is horizon dependence. In both operational and research
verification, simulation quality is interpreted relative to simple reference
distributions such as climatology and persistence. This is directly relevant
for regional temperature simulation. At short horizons, dynamical propagation
of the reduced-rank state can add information beyond climatology. At longer
horizons, that advantage should diminish. This suggests a hybrid formulation in
which the dynamical simulator is regularized toward climatology as
lead time increases \citep{buizza2015fsh,wilks2011statmethods,
vannitsem2021review,hemri2020blending,slater2023hybrid}. This work also fits naturally within statistical climatology that emphasizes
interpretable stochastic generators, reduced-state dynamics, and careful
probabilistic calibration. Our contribution differs from related weather
generators and post-processing methods \citep{bessac2016wind,
gebetsberger2019skewlogistic,gobet2025hmm,baran2025mlsolar} in its focus on a
stable reduced-rank EOF state, explicit subspace-stability diagnostics under
rolling refits, and horizon-adaptive regularization toward climatology for the
reconstructed rank-$K^*$ field. Against this background, we develop an
interpretable low-rank state-space model
for the dominant large-scale component of daily regional temperature fields.
Within each rolling-origin fold, we estimate an EOF basis, retain a compact
subspace, model the retained coefficients through coefficient-specific
seasonal mean and scale functions, propagate the standardized state with a
stable ridge-regularized VARX model, and enrich the innovation specification
only where residual diagnostics indicate that extra structure is needed. The
resulting framework combines reduced-rank representation, coefficient-level
climatological interpretation, stable multivariate dynamics, and
diagnostic-driven stochastic regularization.

We emphasize that the aim is not operational weather forecasting at native
resolution. The model does not seek to resolve fine-scale spatial variability
or compete with numerical weather prediction systems in short-range
deterministic skill. Rather, it provides a statistically transparent framework
for simulating the dominant large-scale regional temperature patterns and for
studying how their predictive structure changes with lead time across
climatically distinct domains.

The contribution of the paper is threefold. First, it treats reduced rank as a
substantive statistical representation of dominant regional temperature
structure, not only as a device for compression. Second, it shows that the
leading retained EOF coefficients admit direct climatological interpretation in
both mean and scale, so that the reduced-rank state is useful for inference as
well as simulation. Third, it shows that simulation quality is strongly
horizon dependent and that a simple horizon-adaptive regularization toward
climatology can stabilize ensemble performance across short-, mid-,
and longer-range lead times.


\section{Related work}
\label{sec:related}

This paper lies at the intersection of stochastic weather generation,
reduced-rank spatio-temporal modeling, multivariate state dynamics,
statistical post-processing, and probabilistic verification. The contribution
is the combination of these components into an interpretable low-rank
state-space simulator for multi-horizon regional temperature fields. Closely
related precedents include regime-switching autoregressive weather generators
\citep{bessac2016wind}, calibration-oriented temperature post-processing
\citep{gebetsberger2019skewlogistic}, interpretable latent-state stochastic
weather generation \citep{gobet2025hmm}, and machine-learning-based
probabilistic calibration studies \citep{baran2025mlsolar}.

\subsection{Weather generators and multisite field simulation}

Classical stochastic weather generators were designed to reproduce the
distributional properties of daily weather series over long periods and remain
important tools in hydrology, agriculture, and climate-impact assessment
\citep{richardson1981weather,semenov1997wg,wilks2011statmethods,
ailliot2015wg_overview}. Their strengths are parsimony, climatological
fidelity, and the ability to generate long synthetic records. The main
difficulty arises when moving from single-site generation to multisite or
gridded simulation, where dependence must be represented coherently across
space and time.

A substantial literature has addressed this problem. Correlated-innovation
weather generators extend classical formulations to multiple locations
\citep{wilks1998multisite,wilks1999multisite}. In forecast post-processing,
methods such as the Schaake shuffle and ensemble copula coupling restore
dependence after marginal calibration and have become standard tools for
producing spatially coherent scenarios \citep{clark2004schaake,
schefzik2013ecc}. Recent work on multisite simulation has also emphasized the
need to represent spatially coherent and compound events
\citep{brunner2021multisite}. \citet{cognot2025tempwg} develop a
spatio-temporal weather generator for temperature over France explicitly
targeting climate-impact assessment including persistent heatwave simulation,
and \citet{nguyen2024nsrwg} present a non-stationary, climate-informed
regional weather generator designed to assess future flood risks, emphasizing
the need for long synthetic meteorological series at regional scales.
\citet{haug2020spatial} show that treating gridded temperature cells
independently, while ignoring spatial coherence, can severely distort trend
assessment, motivating models that respect large-scale spatial structure. Our
objective is closely aligned with this literature, but the modeling strategy
differs: rather than calibrating margins first and reimposing dependence
afterward, we model the dominant large-scale dependence directly through a
reduced-rank latent state.

\subsection{Relationship to deep-learning weather emulators}

Recent years have seen rapid development of deep-learning-based weather
emulators that operate on global fields and can produce multi-day forecasts at
a fraction of the cost of numerical weather prediction
\citep{bi2023pangu,lam2023graphcast,price2024gencast}. These models typically
target deterministic or ensemble prediction at native NWP resolution and are
trained end-to-end on reanalysis data. Our work differs in scope, objective,
and interpretive emphasis. We aim to construct a statistically transparent, low-dimensional simulator for the dominant large-scale component of regional temperature fields over horizons
extending well beyond the deterministic predictability limit. The emphasis on
subspace stability, coefficient-level climatological interpretation, and
diagnostic-driven model building distinguishes the present framework from
data-driven emulators, which prioritize forecast accuracy but typically offer
less interpretive structure. The two approaches are complementary: an emulator
could supply initial conditions or short-range calibration targets, while the
reduced-rank state-space framework developed here provides an interpretable
probabilistic representation for medium- to long-range conditional simulation.

\subsection{Reduced-rank environmental modeling and EOF stability}

High-dimensional environmental fields are often modeled through latent-process
representations, covariance constructions, or basis-function decompositions
\citep{stein1999kriging,gneiting2002nonsep,cressie2011st}. Reduced-rank
approaches provide an attractive compromise between statistical structure and
computational tractability, and underpin methods such as dimension-reduced
state-space filtering and fixed-rank kriging \citep{wikle1999dr_kalman,
cressie2008frk}.

Within atmospheric and climate science, EOFs remain one of the standard tools
for summarizing coherent large-scale variability
\citep{jolliffe2002pca,hannachi2007eofreview}. Their usefulness depends not
only on variance capture, but also on stability. Weak eigenvalue separation can
lead to substantial sampling variability in estimated EOFs
\citep{north1982sampling}, while perturbation theory relates eigenspace
stability to spectral gaps and the size of the underlying perturbation
\citep{davis1970rotation,stewart1990matrix}. This issue is especially relevant
in a rolling-origin design, where the basis is repeatedly re-estimated on
changing training windows. In our setting, the statistical question is not
simply whether EOFs reduce dimension effectively, but whether the retained
subspace is stable enough to support interpretation and conditional simulation.

\subsection{Multivariate dynamics, regularization, and hybridization}

Once the temperature field is projected into EOF space, simulation becomes a
problem in multivariate state dynamics. Vector autoregressions offer a natural
and interpretable framework for representing serial and cross-lag dependence
\citep{hamilton1994ts,lutkepohl2005var}. In moderate dimension, however,
repeatedly refit unrestricted VAR models can be unstable and have high
estimation variance. This motivates shrinkage and regularization, including
ridge-type penalties and Bayesian prior formulations such as the Minnesota
prior \citep{doan1984forecast,litterman1986bvar,banbura2010largebvar}. A useful precedent is the regime-switching autoregressive framework of \citet{bessac2016wind}, which similarly treats stochastic generation through a low-dimensional dynamical representation, though with hidden or observed regimes rather than an EOF-based reduced-rank state.
Related ideas appear in the statistical post-processing literature, where
recent reviews emphasize the need for calibrated methods that remain
computationally tractable while preserving dependence in high-dimensional
forecast distributions \citep{vannitsem2021review,hemri2020blending,
lerch2020multivariate}. Hybrid forecasting provides a particularly relevant
reference point. In that literature, dynamical output is combined with
statistical structure to improve robustness, calibration, or long-range skill
\citep{slater2023hybrid}. Our hybrid construction is deliberately simple: it
does not replace the reduced-rank dynamical simulator, but regularizes it
toward climatology as lead time increases and the incremental value
of dynamical propagation declines.

\subsection{Innovation modeling and probabilistic verification}

Realistic ensemble simulation depends on the innovation specification as well as
on the conditional mean dynamics. Time-varying variance has long been modeled
through ARCH- and GARCH-type recursions \citep{engle1982arch,
bollerslev1986garch}, while dynamic conditional correlation models provide a
classical route to multivariate innovation dependence \citep{engle2002dcc}.
Recurrent neural architectures such as GRUs and LSTMs have also been used for
nonlinear sequential modeling, though in our framework they are used in a much
narrower role: only for residual innovation-scale estimation, not as
end-to-end predictors of the temperature field \citep{hochreiter1997lstm,
cho2014gru}. \citet{schoenach2020profiles} demonstrate that univariate
post-processing of temperature profiles fails to preserve physically
important dependence structure and that copula-based dependence restoration
is needed, a finding that motivates our direct modeling of cross-mode
dependence rather than marginal-then-recouple strategies.
Because the object of interest is an ensemble distribution over fields,
verification must go beyond pointwise error summaries. Proper scoring rules
provide the general framework for probabilistic evaluation
\citep{hersbach2000crps,dawid1984prequential,diebold1998density}. For
temperature post-processing in particular,
\citet{gebetsberger2019skewlogistic} provide a useful reference for the
calibration-versus-sharpness perspective and for PIT-based assessment in a
temperature setting. For high-dimensional spatial quantities, however,
dependence matters directly, and multivariate scores such as the energy score
and variogram score have become standard for ensemble-field verification
\citep{gneiting2007scoring,scheuerer2015variogram}. \citet{pic2025scoring}
argue that multivariate verification should use multiple interpretable proper
scoring rules rather than a single summary, a perspective we adopt by reporting
energy score, variogram score, and a warm-exceedance area CRPS diagnostic
jointly. These criteria are well suited to our goal of producing ensembles
whose spatial coherence, persistence, and calibration remain credible across
lead times.

\section{Data and experimental protocol}
\label{sec:exp}

\subsection{Temperature data}
\label{sec:data_primary}

We analyze daily regional temperature fields derived from hourly 2\,m air
temperature in the ERA5 global reanalysis, produced by ECMWF and distributed
through the Copernicus Climate Data Store \citep{hersbach2020era5,
cds_era5_single_levels}. For each grid cell, the daily value is defined as the
mean of the 24 hourly 2\,m temperatures for that day. ERA5 provides a
spatially complete and physically consistent reconstruction and is widely used
in regional climate analysis and model evaluation. In this paper, it serves as
the reference field whose dominant large-scale daily variability we seek to
represent statistically.

The study period extends from 1 January 1940 to 31 December 2025. This long
record spans multiple decades of forced change and internal variability and
therefore provides a stringent setting for assessing subspace stability,
parameter robustness, and multi-horizon simulation performance.

\subsection{Regional domains}
\label{sec:domains}

We consider five climatically distinct U.S.\ regional domains extracted from
ERA5 on the native $0.25^\circ$ latitude-longitude grid. Each domain is
defined by a fixed bounding box, aligned to the ERA5 grid and held fixed over
time and across all rolling-origin folds. Because the test periods overlap in calendar time, cross-fold variability should be interpreted as a robustness summary rather than as independent-sample variation. This is consistent with the broader time-series forecast-evaluation literature, which emphasizes that dependence and overlap affect how cross-validation or rolling-origin comparisons should be interpreted \citep{bergmeir2018cv}. For each region, the daily field is
represented on a fixed $64\times64$ grid, so that each domain spans
approximately $15.75^\circ$ in each coordinate direction.

The domains were selected to reflect contrasting climatic and physiographic
settings, including coastal influence, continentality, and complex terrain.
This diversity allows us to assess whether a common reduced-rank modeling
strategy remains effective across markedly different regional environments. The
five domains (Fig.~\ref{fig:eof_maps}) are:
\begin{itemize}
\item \textbf{Upper Midwest:}
North 49.75, West $-95.50$, South 34.00, East $-79.75$.
\item \textbf{Southern California:}
North 41.75, West $-126.25$, South 26.00, East $-110.50$.
\item \textbf{Gulf-South Central:}
North 37.50, West $-103.25$, South 21.75, East $-87.50$.
\item \textbf{Central Rockies:}
North 47.50, West $-113.00$, South 31.75, East $-97.25$.
\item \textbf{Northeast Corridor:}
North 48.50, West $-82.00$, South 32.75, East $-66.25$.
\end{itemize}

The compact label ``Southern California'' is retained for consistency across the
analysis products; its fixed box intentionally includes the broader
California--Southwest domain and adjacent Pacific grid cells.

At this scale, \emph{regional} refers to coherent mesoscale-to-synoptic
structure rather than local pointwise variation. This matches the scope of the
model, which is intended to represent dominant large-scale spatial variability.
It is not designed to resolve sub-grid structure or to serve as a local
forecasting tool.

\subsection{Calendar features and external covariates}
\label{sec:covariates}

The model conditions only on information available at the simulation origin. We
therefore distinguish between deterministic calendar features, which are known
in advance, and optional low-frequency external covariates, which summarize
broader climate variability.

\subsubsection{Calendar features}

Let $\mathrm{doy}(t)\in\{1,\ldots,366\}$ denote day of year, and define
$\theta_t=\mathrm{doy}(t)/365.25$. This fixed normalization is the convention
used in the reported analysis; its very small leap-year boundary mismatch is
negligible relative to the smooth harmonic representation. Seasonal structure
is represented through $H$ harmonic pairs,
\[
h(\theta_t)=
\bigl(\sin(2\pi j\theta_t),\cos(2\pi j\theta_t)\bigr)_{j=1}^H,
\]
together with an optional deterministic drift term based on normalized time.

\subsubsection{Low-frequency climate indices}

To allow the coefficient-level mean and scale structure to respond to slowly
varying external forcing and low-frequency climate variability, we include
three monthly indices:
\begin{itemize}
\item \textbf{CO$_2$:} monthly mean atmospheric CO$_2$ from the NOAA Global
Monitoring Laboratory \citep{noaa_gml_co2_data};
\item \textbf{ENSO (ONI):} the Oceanic Ni\~no Index from NOAA Climate
Prediction Center \citep[Version~5; values through the 2025 study
endpoint]{noaa_cpc_oni_data};
\item \textbf{AMO:} the Atlantic Multidecadal Oscillation index from NOAA PSL
\citep[Kaplan-SST series through January~2023]{noaa_psl_amo_data}.
\end{itemize}
These indices are used as descriptive low-frequency controls in coefficient
space. Because CO$_2$, ONI, AMO, and screened slow drift are correlated over
the historical record, we interpret their fitted coefficients as conditional
statistical associations rather than causal effects.

The monthly source records do not have identical native temporal coverage over
the full 1940--2025 ERA5 study period. Appendix
Table~\ref{tab:covariate_coverage} reports the native coverage and the exact
leading or trailing intervals filled for each index. In the implementation, the
monthly records are first combined into a single covariate table, aggregated to
month-start timestamps, and then filled by forward fill; any leading missing
values are back-filled with the first available monthly value. This convention
is applied before training-only standardization. Consequently, parts of the
study record contain constant extrapolated values where an index is not
natively observed. This is another reason these low-frequency variables are
interpreted only as descriptive controls rather than as causal forcing
estimates.

Because the climate indices are observed monthly and evolve slowly, they are
treated as fixed at their most recently available origin-time values over the
simulation horizon. For the horizons considered here, this is a reasonable
approximation for CO$_2$ and AMO on sub-annual scales. ONI can evolve
appreciably over 90--120 days, particularly during boreal spring transitions,
so the fixed-covariate approximation is strongest at short and medium leads and
should be regarded as progressively weaker beyond roughly 60 days.

\subsubsection{Temporal alignment}

The external indices are observed monthly. We first align the monthly series to
the daily ERA5 dates by backward as-of matching: each daily date receives the
most recent month-start value available at or before that date. We then shift
the resulting daily covariate series by a fixed causal lag,
\[
x_t = x^{\mathrm{asof}}_{t-\ell},
\]
with $\ell=1$~day in the reported experiments. Thus \(x_t\) is the one-day
lagged, as-of-aligned monthly index. This construction prevents the use of
future monthly timestamps, but it should not be interpreted as a strict
previous-calendar-month assignment for every day in a month.

\section{Methodology}
\label{sec:method}

\subsection{Goal and scope}
\label{sec:method:goal}

The goal is to construct a statistically interpretable simulator for the
retained EOF-projected temperature field. The model operates in a
low-dimensional coefficient space and reconstructs simulated trajectories to
grid space through the retained EOF basis.

\subsection{Overview and state-space interpretation}
\label{sec:method:overview}

Let $Y_t \in \mathbb{R}^D$ denote the daily gridded temperature field over a
fixed regional domain, with $D=4096$ in our applications. Because most of the
large-scale variability is concentrated in a small number of coherent spatial
modes, we represent the field in empirical orthogonal function (EOF) space and
treat the retained EOF coefficient vector as a low-dimensional state. Mapping
through the retained EOF basis returns the state to physical space.

Within each rolling-origin fold, the model has four components:
\begin{enumerate}
\item a fold-specific EOF representation of the daily field;
\item coefficient-specific deterministic mean and scale functions for the
retained EOF amplitudes;
\item a stable multivariate dependence model for the standardized retained
coefficients;
\item a structured innovation specification for the remaining uncertainty,
together with an optional horizon-adaptive regularization toward
climatology.
\end{enumerate}

The model does not target the full native-rank field directly. Rather, it
targets the dominant retained low-rank component. In the reported experiments,
simulation stops at that reconstructed rank-$K^*$ field; no additional
native-space residual correction is added.

\begin{figure}[t]
\centering
\includegraphics[width=0.92\linewidth]{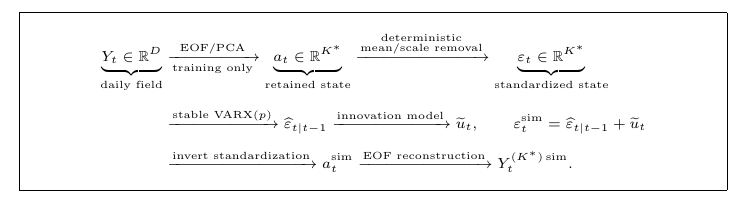}
\caption{Construction within one rolling-origin fold. The daily field is
projected onto a training-estimated EOF subspace, deterministic structure is
removed in coefficient space, the standardized reduced-rank state is propagated
through stable multivariate dynamics, and the remaining uncertainty is modeled
through a calibrated innovation specification.}
\label{fig:method_flow}
\end{figure}

Figure~\ref{fig:method_flow} summarizes the construction. The reduced-rank
state is estimated using training data only. Deterministic seasonal structure,
optional drift, and optional low-frequency covariate effects are removed
coefficient by coefficient. The resulting standardized state is propagated by a
regularized VARX model under explicit stability control. Residual uncertainty
is then represented through an innovation specification that allows
time-varying marginal scale, contemporaneous cross-mode dependence, and
dispersion calibration, all estimated from training data only. Simulated
trajectories in coefficient space are finally mapped back to the grid through
the retained EOF basis.

\subsection{Predictive target}
\label{sec:target}

For simulation origin $o$ and lead time $h\ge1$, let
\[
\mathcal{F}_o=\sigma\!\left(\{Y_t,x_t\}_{t\le o}\right),
\]
where $x_t$ denotes covariates available at time $t$ without look-ahead.

The inferential target is the projected predictive law
\[
\mathcal{L}\!\left(Y_{o+h}^{(K^*)}\mid\mathcal{F}_o\right),
\]
where $Y_t^{(K^*)}$ denotes the rank-$K^*$ projection of $Y_t$ onto the
fold-specific EOF subspace estimated from training data. Thus, the model
targets the dominant large-scale component of the regional field rather than
the full native-rank field. In practice, the fitted model yields an ensemble
approximation to this projected predictive distribution by simulating
trajectories in EOF coefficient space and reconstructing them through the
retained EOF basis.

This distinction is central. Variability outside the retained EOF subspace is
not modeled explicitly. The adequacy of the truncation is therefore evaluated
empirically through explained variance, eigenvalue separation, subspace
stability, residual temporal dependence after truncation, and out-of-sample
verification of the reconstructed low-rank field.

\subsection{EOF representation}
\label{sec:eof_method}

Fix a rolling-origin training fold and let $\mathcal{I}_{\mathrm{tr}}$ denote
its training index set. Define the training mean field
\[
\mu_{\mathrm{tr}}
=
\frac{1}{|\mathcal{I}_{\mathrm{tr}}|}
\sum_{t\in\mathcal{I}_{\mathrm{tr}}} Y_t,
\]
and centered training fields
\[
X_t = Y_t - \mu_{\mathrm{tr}}.
\]

The sample covariance matrix of the centered training fields is
\[
\widehat{\Sigma}_{\mathrm{tr}}
=
\frac{1}{|\mathcal{I}_{\mathrm{tr}}|-1}
\sum_{t\in\mathcal{I}_{\mathrm{tr}}} X_tX_t^\top
=
V\Lambda V^\top,
\]
where $V=(v_1,\ldots,v_D)$ contains orthonormal EOF loadings and
$\Lambda=\mathrm{diag}(\lambda_1,\ldots,\lambda_D)$ contains the associated
eigenvalues in decreasing order. For any $K\le D$, let
$V_K=(v_1,\ldots,v_K)\in\mathbb{R}^{D\times K}$ denote the retained basis. The
corresponding coefficient vector is
\[
a_t^{(K)} = V_K^\top X_t,
\]
and the rank-$K$ reconstruction is
\[
Y_t^{(K)} = \mu_{\mathrm{tr}} + V_K a_t^{(K)}.
\]

All EOF calculations in the reported experiments use equal grid-cell weights on
the fixed \(64\times64\) regional grid. The resulting target is therefore the
equal-grid-cell projected field, rather than an area-weighted hemispheric or
global integral. This choice is appropriate here because each domain is a fixed
regional box and all models, scores, and diagnostics are applied consistently
to the same projected grid representation.

In the reported experiments we retain \(K^*=4\) EOF coefficients to capture at
least 95\% of the training variance while keeping the state dimension small
enough for stable multivariate dynamics and direct coefficient-level
interpretation:
\[
\frac{\sum_{j=1}^{4}\lambda_j}{\sum_{j=1}^{D}\lambda_j} > 0.95.
\]
Across the five regions and rolling folds, the retained four-mode state explains
about 96.7--97.5\% of centered-field variance. These fractions are for centered
daily temperature fields, not for deseasonalized anomaly fields. This is
intentional: the target of the present generator is the dominant daily
regional temperature field, and the annual cycle is one of the most important
sources of structure in that field. Seasonal low-rank structure is therefore
part of the state to be represented, after which its deterministic mean and
scale modulation are modeled explicitly in coefficient space. Thus all
simulations and diagnostics are conducted for the reconstructed rank-\(4\) field
\(Y_t^{(K^*)}\). The retained rank is chosen for the generator because it
captures the dominant thermal mode together with the leading regional contrast
modes while preserving a compact, interpretable, and dynamically stable state.

Reduced-rank representations of environmental fields are well established. The
critical question for the present framework is not whether EOFs reduce
dimension effectively--they plainly do--but whether the retained \(K^*=4\)
subspace is stable enough under repeated rolling refits to support a persistent
statistical interpretation of the leading coefficients and to justify
rolling-origin simulation in the reduced state.

\begin{figure}[t]
\centering
\includegraphics[width=\textwidth]{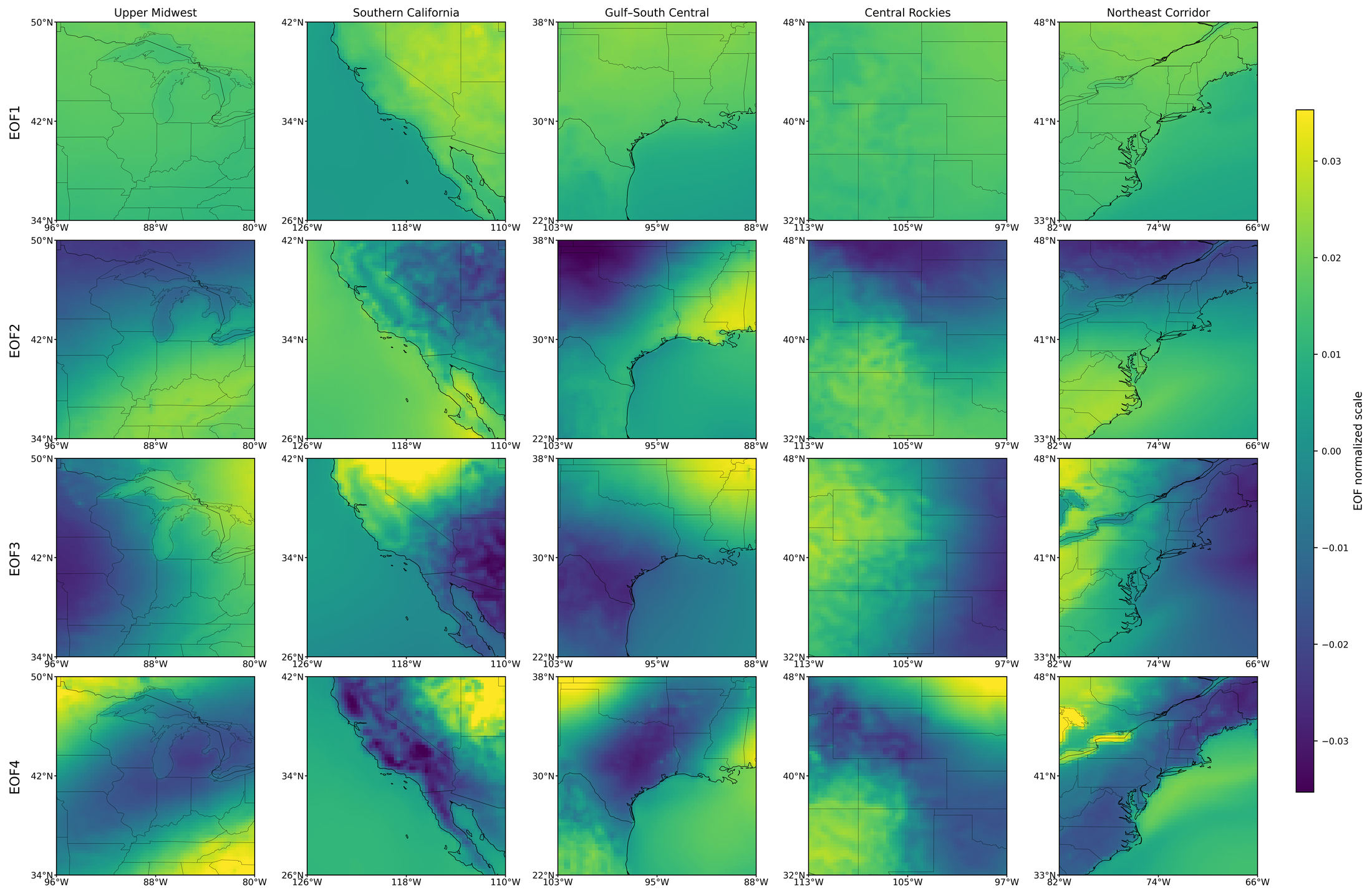}
\caption{Leading EOF loading patterns across the five regional domains for the
retained \(K^*=4\) state. EOF~1 is a spatially weighted domain-scale thermal
mode, with loadings of common sign across each grid, so that its coefficient
describes a broadly coherent warm-versus-cold regional fluctuation. EOFs~2--4
represent lower-variance regional contrast modes, including north--south,
coast--interior, terrain-related, and ocean--land structures depending on the
domain. The loading maps are spatially coherent across all five regions,
supporting the interpretation of the retained coefficients as amplitudes of
large-scale regional temperature modes rather than arbitrary basis
coordinates.}
\label{fig:eof_maps}
\end{figure}

\subsection{Deterministic structure in coefficient space}
\label{sec:coef_decomp_interp}

Let $a_{t,k}$ denote retained EOF coefficient $k$, for
$k=1,\ldots,K^*$. We write
\begin{equation}
a_{t,k}=m_k(t)+s_k(t)\varepsilon_{t,k},
\label{eq:coef_decomp_final}
\end{equation}
where $m_k(t)$ is a deterministic mean function, $s_k(t)>0$ is a deterministic
scale function, and $\varepsilon_{t,k}$ is the standardized residual
coefficient.

Our working specification is
\begin{align}
m_k(t)
&=
\alpha_{k,0}
+\alpha_k^\top h(\theta_t)
+\beta_k^\top x_t
+\delta_k t,
\label{eq:mk_final}\\
\log s_k^2(t)
&=
\gamma_{k,0}
+\gamma_k^\top h(\theta_t)
+\eta_k^\top x_t,
\label{eq:sk_final}
\end{align}
where $\theta_t$ is normalized day of year, $h(\theta_t)$ is a harmonic basis
for the annual cycle, $x_t$ denotes optional low-frequency covariates available
without look-ahead, and $t$ is a normalized time index used to capture slow
drift. The drift term is included as a parsimonious descriptive adjustment for
slow nonstationarity. In the implementation, drift is screened only for
PCs\,1--3 and is included only when a Theil--Sen screening slope
\citep{theil1950rank,sen1968estimates} has a confidence interval excluding
zero. For PC 1, this screening slope is positive
whenever drift is selected: it is selected in 3/4 folds for Upper Midwest, 4/4 for
Central Rockies, 3/4 for Gulf--South Central, 4/4 for Southern California, and 2/4 for Northeast Corridor. This is
consistent with PC 1 acting as a domain-scale thermal mode. For higher PCs,
the selected drift terms can change sign across regions, reflecting changes in
regional contrasts rather than uniform warming. Because CO$_2$ and other
low-frequency covariates are collinear over the historical record, we
interpret individual drift and covariate coefficients descriptively rather than
causally. The drift estimates are therefore most useful as a diagnostic that
the leading EOF coefficient behaves like a warming-sensitive domain-scale
thermal amplitude, not as a separate attribution of forced warming.

The standardized state is then
\[
\varepsilon_{t,k}
=
\frac{a_{t,k}-m_k(t)}{s_k(t)},
\qquad
\varepsilon_t=(\varepsilon_{t,1},\ldots,\varepsilon_{t,K^*})^\top.
\]

This decomposition gives the retained low-rank state a direct climatological
interpretation. The fitted mean and scale functions describe how the amplitudes
of the dominant regional modes vary over the annual cycle and respond to slow
drift or low-frequency covariates. The reduced-rank state is therefore useful
for inference as well as simulation.

\subsection{Dependence model for the standardized state}
\label{sec:varx}

Joint temporal dependence in the standardized state is modeled by a
ridge-regularized VARX($p$) system,
\begin{equation}
\varepsilon_t
=
c+\sum_{i=1}^{p}A_i\varepsilon_{t-i}+B\widetilde{x}_t+u_t,
\label{eq:varx_full}
\end{equation}
where $\widetilde{x}_t$ denotes exogenous predictors available at time $t$
without look-ahead and $u_t$ is the one-step innovation vector.

The lag order $p$ is selected using training data only from a finite candidate
set, and estimation is regularized by ridge penalization. Because the fitted
system is used for iterated multi-step simulation, dynamic stability is
essential; only fits whose companion matrix satisfies the preset stability
criterion are retained.

To discourage excess persistence in higher-order coefficients under iterated
simulation, the ridge target for the VARX is set proportional to the inverse
marginal variance of each coefficient, with a regularization strength
parameter $\lambda_{\mathrm{target}}$. This ensures that lower-variance
higher-order modes receive proportionally stronger shrinkage toward zero
dynamics, while the dominant mode retains more freedom to capture its stronger
predictable serial dependence. In the final implementation, a descending mode-specific lag cap is used:
PC~1 uses the selected order \(p\), and retained mode \(k\) uses
\(p_k=\max\{p-k+1,1\}\). Thus each successive mode is allowed one fewer
lag, down to a minimum of one. This is a regularization heuristic designed to
prevent lower-variance contrast modes from introducing excessive long-lag
persistence under iterative simulation; it is not intended as a claim that
higher-order empirical autocorrelation is strictly ordered in every region.

This stage captures predictable linear dependence in the retained state,
including serial dependence within coefficients and cross-lag dependence across
coefficients. Such cross-lag effects matter because the leading EOF amplitudes
typically evolve on different persistence scales and interact dynamically.

\begin{figure}[t]
\centering
\includegraphics[width=\textwidth]{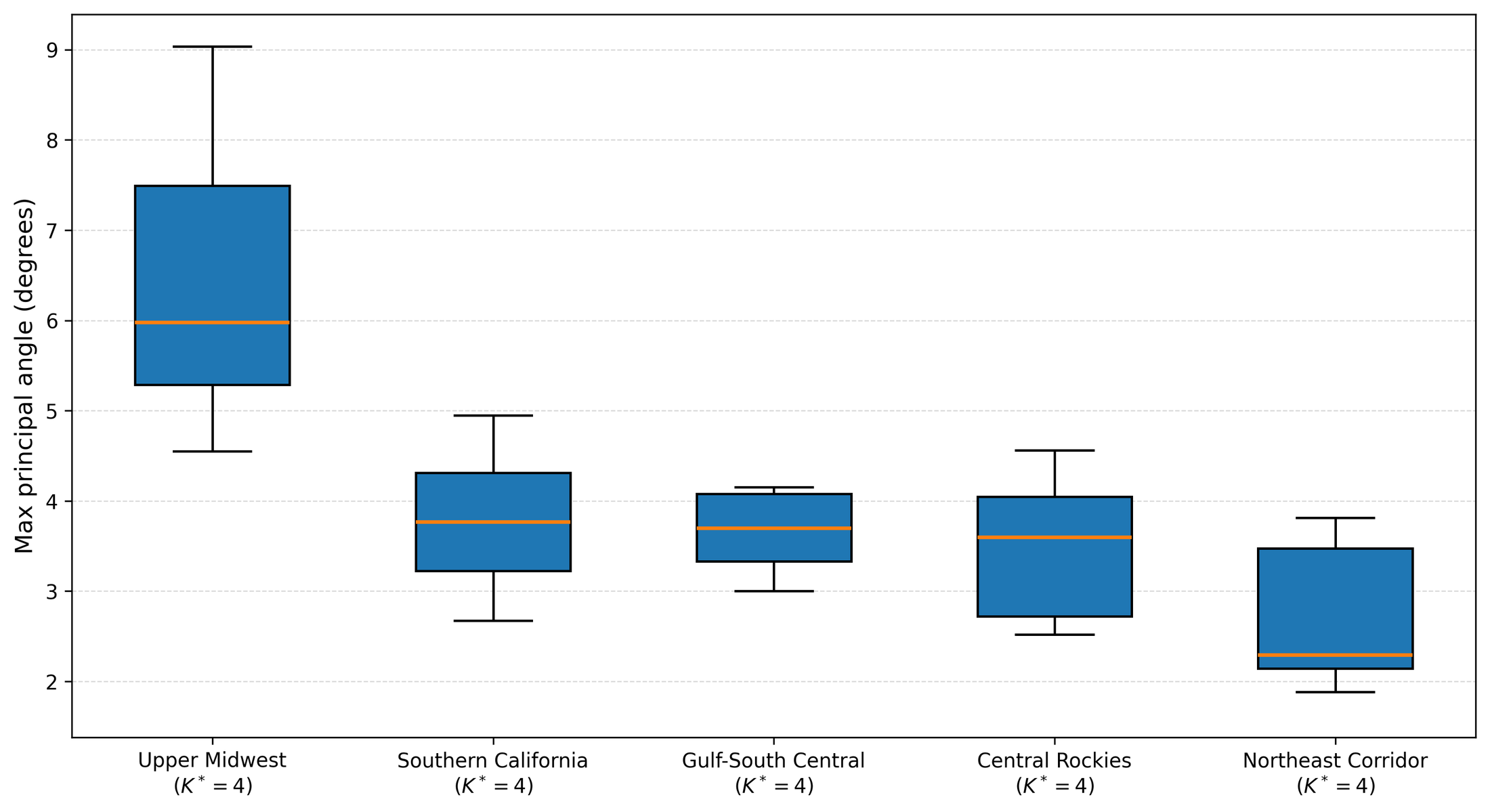}
\caption{Principal-angle stability of the retained EOF subspace across regions.
The boxplots show the maximum principal angle (in degrees) between the
full-training $K^*$-dimensional subspace and rolling 20-year sub-block refits.
Smaller angles indicate greater stability of the dominant large-scale
structure.}
\label{fig:stab_max_angle}
\end{figure}

\subsection{Model classes used in the reported experiments}
\label{sec:model_family}

We compare one climatological benchmark, three dynamical models,
and their corresponding hybrids.

\begin{itemize}
\item \textbf{CLIM}: Conditional climatology obtained by resampling contiguous
training episodes with similar calendar context. For a simulation origin with
day-of-year \(d\), CLIM draws contiguous multi-day segments from the training
record whose starting day-of-year lies within a window \([d-15,d+15]\).
Because episodes are sampled as blocks, they retain the temporal dependence
structure of the training record. When the initial candidate pool contains
fewer than 250 start dates, the window is expanded in steps of 5 days up to a
maximum of \(\pm 60\) days.

\item \textbf{$\mathrm{M}_1$}: baseline dynamical model with constant
innovation covariance estimated from training residuals. It propagates the
standardized state through the fitted VARX model and uses the full Cholesky
factor of the residual covariance $\widehat{\Sigma}_u$ together with
Student-t shocks. A scalar dispersion calibration is then applied to the M1
innovations, with the scalar constrained to the interval \([0.90,1.50]\).

\item \textbf{$\mathrm{M}_2$}: heteroskedastic dynamical model based on an
exponentially weighted moving average (EWMA) recursion \citep{riskmetrics1996}
for marginal innovation scale. The smoothing parameter $\lambda$ is tuned on an
inner training/validation split over the grid
$\{0.90,0.93,0.95,\allowbreak 0.97,0.98,0.99\}$. $\mathrm{M}_2$ uses season-specific cross-mode
dependence and componentwise dispersion calibration.

\item \textbf{$\mathrm{M}_3$}: heteroskedastic dynamical model in which a GRU
provides a time-varying volatility correction in coefficient space. In our
implementation, the GRU scale is not used independently of the EWMA scale.
Instead, the GRU-to-EWMA ratio is constrained to lie in $[c^{-1},c]$ with
$c=2$, and the constrained GRU estimate is then geometrically averaged with
the EWMA baseline using blend weight 0.5. This regularization prevents unstable
scale paths during iterated simulation while allowing $\mathrm{M}_3$ to depart
meaningfully from pure EWMA behavior. As in $\mathrm{M}_2$,
$\mathrm{M}_3$ uses season-specific cross-mode dependence and multi-level
dispersion calibration.

\item \textbf{Hybrid-$\mathrm{M}_1$}, \textbf{Hybrid-$\mathrm{M}_2$},
\textbf{Hybrid-$\mathrm{M}_3$}: trajectory-level hybrids that follow the
corresponding dynamical simulator at short leads and then stochastically branch
to state-conditioned climatological continuations as lead time increases.
\end{itemize}

All reported simulations use the same deterministic mean-scale decomposition,
the same stable VARX dependence model, and Student-t innovation shocks. The
models differ in the structure and calibration of the remaining innovation
uncertainty: M1 uses a constant covariance with scalar dispersion calibration,
M2 uses EWMA marginal volatility with componentwise calibration, and M3 uses an
EWMA-regularized GRU volatility correction with the same dependence and
calibration framework as M2.

\subsection{Innovation specification}
\label{sec:innovation}

After deterministic normalization and VARX propagation, the remaining
uncertainty is represented by the one-step residual
\[
\widehat{u}_t
=
\varepsilon_t-\widehat{c}
-\sum_{i=1}^{p}\widehat{A}_i\varepsilon_{t-i}
-\widehat{B}\widetilde{x}_t.
\]

For $\mathrm{M}_1$, innovations are generated from the constant training
residual covariance:
\[
\widetilde{u}_t^{(\mathrm{M1})}
=
L_u z_t,
\]
where \(L_u\) is the Cholesky factor of \(\widehat{\Sigma}_u\) and \(z_t\) is
a standardized Student-t shock vector with per-component degrees of
freedom. The simulated M1 innovation is then
\[
u_t^{(\mathrm{M1})}
=
\gamma_{\mathrm{M1}}\widetilde{u}_t^{(\mathrm{M1})},
\]
where \(\gamma_{\mathrm{M1}}\) is a scalar dispersion-calibration factor
estimated from training residual diagnostics and constrained to
\([0.90,1.50]\).

For $\mathrm{M}_2$, the marginal scale path follows the EWMA recursion
\[
\sigma^2_{t,k}
=
\lambda\,\sigma^2_{t-1,k}
+
(1-\lambda)\,\widehat{u}^2_{t-1,k},
\]
with $\lambda$ tuned from training data over the grid
\(\{0.90,0.93,0.95,0.97,0.98,0.99\}\). For $\mathrm{M}_3$, the GRU provides a
time-varying volatility correction that is regularized toward the EWMA
baseline as described in Sect.~\ref{sec:model_family}. In the reported
experiments, a small lead-dependent inflation is applied only to the M3
innovation scale at the first three simulated leads, with multipliers 1.10,
1.05, and 1.02; no additional short-horizon inflation is applied beyond lead 3.

For the heteroskedastic models (\(\mathrm{M}_2\) and M3), a
coefficient-specific innovation-persistence blend is applied before final
dispersion calibration:
\begin{equation}
\breve{u}_{t,k}
=
\rho_k\,\breve{u}_{t-1,k}
+
\sqrt{1-\rho_k^2}\,\widetilde{u}_{t,k},
\qquad k=1,\ldots,K^*.
\label{eq:persist_pc}
\end{equation}
In the reported experiments, \(\boldsymbol{\rho}=(0.1,0,0,0)^\top\).
Thus a mild persistence stabilization is applied only to the dominant mode,
while higher-order coefficients receive no additional innovation persistence.
This asymmetric specification is motivated by the empirical persistence
structure of the retained state: PC 1 is consistently the most persistent
coefficient, whereas PCs\,2--4 have weaker and more region-specific
persistence (Table~\ref{tab:m2_acf}). Higher-order persistence is controlled
primarily through the VARX target regularization described in
Sect.~\ref{sec:varx}.

The distinction between \(\sigma_t\) and \(\gamma\) is important. The role of
\(\sigma_t\) is dynamic: it captures time variation in innovation scale as a
function of recent residual history. The role of \(\gamma\) is calibration: it
adjusts residual componentwise dispersion remaining after the scale path has
been estimated. Thus \(\gamma\) is not a second volatility model, but a
post-calibration factor.

The innovation specification should therefore be interpreted narrowly. It does
not replace the deterministic and linear dynamical components; rather, it
models the residual uncertainty that remains after those components have
removed the dominant predictable structure. More broadly, the regularization
devices in the pipeline---the VARX ridge penalty and lag cap, the EWMA-to-GRU
ratio clip, the M3 short-horizon scale inflation, the persistence blend, and
the dispersion-calibration bounds---serve a common purpose: they are
training-only stabilization choices designed to keep the iterated dynamical
system well behaved over multi-step simulation horizons.

\subsection{Climatology and trajectory-branching hybrid}
\label{sec:hybrid}

The quality of conditional simulation is strongly horizon dependent. At short
leads, dynamical propagation of the reduced-rank state adds useful information
beyond climatology. At longer leads, however, the predictive distribution
should increasingly resemble a seasonally conditioned reference. We therefore
combine the dynamical simulator with a climatology baseline using a
trajectory-level branching construction that preserves temporal coherence
within each ensemble member.

\subsubsection{Stochastic trajectory branching}
The CLIM baseline is defined in Sect.~\ref{sec:model_family}. For each ensemble
member~$m$, the hybrid proceeds as follows:
\begin{enumerate}
\item A stochastic branch lead $B_m$ is sampled from a shifted geometric
distribution. Its survival probability is one through the minimum branch lead
and then decays approximately as
$\exp\{-(h-h_{\min})/\tau\}$, where $\tau>0$ is the fitted
skill-horizon time scale. A minimum lead of $h_{\min}$~days is imposed before
any branching occurs.
\item For $h\le B_m$, the trajectory follows the full dynamical model. Thus
$B_m$ is the last purely dynamical lead for member~$m$.
\item For leads $h>B_m$, the trajectory is handed off to a
state-conditioned climatological continuation. The standardized dynamical
state $\varepsilon_{B_m}^{(\mathrm{dyn})}$ at the handoff point is used to
select a donor episode from the training record by nearest-neighbor matching in
the $K^*$-dimensional coefficient space.
\end{enumerate}

\begin{figure}[t]
\centering
\includegraphics[width=0.92\linewidth]{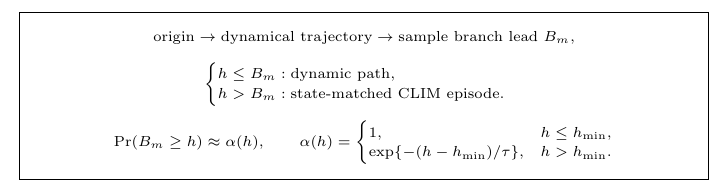}
\caption{Trajectory-level hybridization. Each ensemble member follows the
corresponding dynamical simulator until a random branch lead and then switches
to a state-matched climatological continuation. The ensemble margin therefore
moves toward climatology with increasing lead time, while each member remains a
coherent trajectory rather than a pointwise mixture.}
\label{fig:hybrid_flow}
\end{figure}

The donor selection proceeds within the same day-of-year window used by CLIM.
Among the $n_{\mathrm{top}}$ nearest neighbors, measured by Euclidean distance
in the standardized state, the donor is drawn with probability proportional to
\[
\exp\left(
-\|\varepsilon_{B_m}^{(\mathrm{dyn})}-\varepsilon_{t^*}\|/T
\right),
\]
where $t^*$ is the selected donor time and $T$ is a softmax temperature
parameter. This provides moderate state dependence while retaining stochastic
diversity across ensemble members.

\subsubsection{Offset decay and ramp}
At the handoff point, the dynamical state and selected donor state may differ
slightly even after nearest-neighbor matching. Define the state mismatch as
\(\Delta_m=\varepsilon_{B_m}^{(\mathrm{dyn})}-\varepsilon_{t^*}\).
For the post-branch lead adjustment \(j\ge 1\), the offset-adjusted donor
continuation is
\[
\widetilde{\varepsilon}_{B_m+j}
=
\varepsilon_{t^*+j}
+
\left(1-\frac{j}{\delta}\right)_+ \Delta_m,
\qquad (x)_+ = \max(x,0).
\]
Thus the offset is explicitly down-weighted: at the first post-branch step the
correction is \((1-1/\delta)\Delta_m\), and it vanishes for \(j\ge\delta\).

To avoid an abrupt switch from the dynamical trajectory to the donor
continuation, the first \(R\) post-branch days are linearly ramped. For
\(j=1,\ldots,R\), define
\[
\varepsilon_{B_m+j}^{(\mathrm{hyb})}
=
\beta_j \varepsilon_{B_m+j}^{(\mathrm{dyn})}
+
(1-\beta_j)\widetilde{\varepsilon}_{B_m+j},
\qquad
\beta_j
=
1-\frac{j}{R+1}.
\]
For \(j>R\), we set
\(\varepsilon_{B_m+j}^{(\mathrm{hyb})}=\widetilde{\varepsilon}_{B_m+j}\).
The linear ramp is the natural convex combination for smoothing the conditional
mean at the splice, but it can damp marginal variation during the transition
because the two paths are averaged. We retain it to avoid discontinuities and
account for this limitation when describing the fixed-lead predictive margin.
In the reported experiments, \(h_{\min}=7\), \(R=5\), \(\delta=18\),
\(n_{\mathrm{top}}=3\), and \(T=1.5\).

\subsubsection{Approximate marginal mixture}

At any fixed lead~\(h\), the fraction of ensemble members whose branch lead is
at least \(h\) is approximately
\[
\alpha(h)=
\begin{cases}
1, & h\le h_{\min},\\
\exp\{-(h-h_{\min})/\tau\}, & h>h_{\min},
\end{cases}
\]
up to finite-horizon truncation and the discrete geometric sampling
approximation. Hence, for leads beyond the minimum branch lead, the marginal
predictive distribution is close to the mixture
\[
\mathcal P_h^{(\mathrm{Hybrid}\text{-}\mathcal M)}
\approx
\alpha(h)\,\mathcal P_h^{(\mathcal M)}
+
\bigl(1-\alpha(h)\bigr)\mathcal P_h^{(\mathrm{CLIM})}.
\]
Outside the \(R\)-day ramp, this expression has the same fixed-lead mixture
interpretation as a distribution-level blend. During the ramp, however, the
margin also contains convexly interpolated states and can therefore be slightly
underdispersed. The key difference from a distribution-level blend is that each
trajectory is internally coherent: before the branch it follows the dynamical
simulator, and after the branch it follows a state-conditioned historical
continuation with a smoothed transition. This trajectory-level construction shares motivation with
the SimSchaake approach to dependence reconstruction
\citep{schefzik2016simschaake,scheuerer2017schaake}, which similarly
conditions on atmospheric state when selecting historical dependence
templates, and with the analog ensemble framework
\citep{dellemonache2013anen}, which generates forecasts from state-similar
historical episodes. The difference is that our construction operates within
an iterated low-rank state-space simulator rather than as a post-processing
step.

\subsubsection{Skill-horizon estimation}
The shrinkage time scale $\tau$ is estimated from training data only, using
24~day-of-year bins with smoothness regularization and fitting restricted to
horizons $h\ge 7$. This yields a seasonally varying estimate of the lead time
at which dynamical skill decays to climatological levels, consistent with the
forecast skill horizon concept of \citet{buizza2015fsh}. The regional
differences in $\tau$ (Table~\ref{tab:m9_hybrid_tau}) reflect differences in
the persistence and predictability of the retained temperature modes across
domains.

\begin{figure}[t]
\centering
\includegraphics[width=\textwidth]{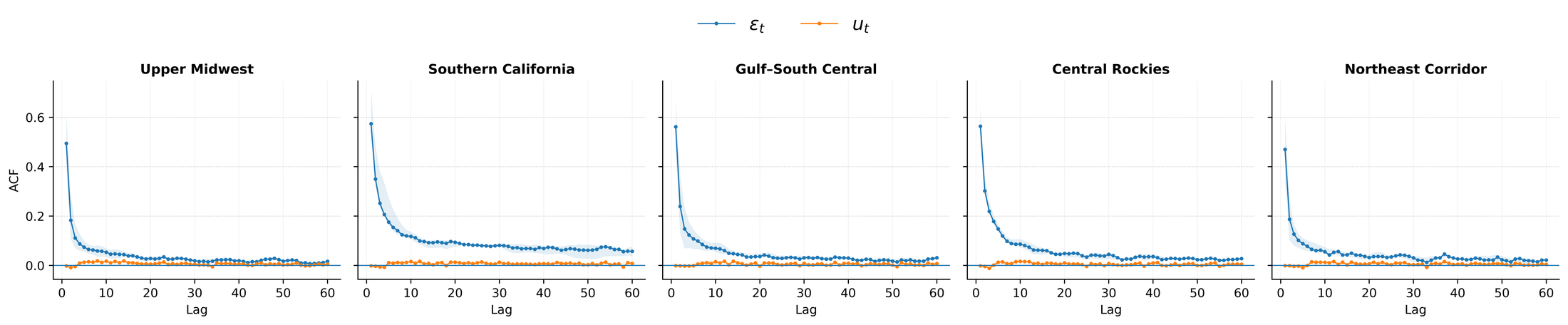}
\caption{Average autocorrelation functions of the standardized coefficient
process $\varepsilon_t$ and the fitted one-step residuals $\widehat u_t$. The
VARX layer removes a substantial fraction of the serial dependence present in
the standardized retained coefficients, while the remaining scale variation
motivates the structured innovation specification.}
\label{fig:acf_train}
\end{figure}

\section{Evaluation framework}
\label{sec:eval}

The evaluation has two distinct goals. First, we assess whether the retained
EOF representation yields a statistically meaningful low-rank state. Second, we
assess whether the resulting rolling-origin ensembles are calibrated,
dependence-aware, and dynamically credible out of sample.

All quantities are estimated using training data only within each fold and then
held fixed out of sample. Because the test folds overlap in calendar time,
pooled cross-fold comparisons are interpreted primarily as robustness summaries
rather than as formal independent-sample significance analyses.

\subsection{Structural checks for the retained low-rank state}
\label{sec:eval:structure}

To assess whether the retained EOF state is more than a compression device, we
report the explained variance of the retained rank $K^*$, the stability of the
retained subspace under rolling refits, and the residual temporal dependence in
the grid field after removing the retained reconstruction. These diagnostics
address the central structural question of the paper: whether the retained EOF
coefficients may be interpreted as amplitudes of a stable, meaningful
large-scale state rather than as window-specific basis coordinates.

\subsection{Out-of-sample ensemble verification}
\label{sec:eval:forecast}

At each simulation origin $o$, the fitted model generates an ensemble
\[
\{Y_{o+h}^{(K^*)\,(m)}\}_{m=1}^{M},
\qquad h\in\mathcal H,
\]
which is compared with the realized projected field $Y_{o+h}^{(K^*)}$.

Our primary verification summaries are as follows. \emph{Calibration} is
assessed using empirical coverage and probability integral transform (PIT)
diagnostics. For a given pixel and lead time, the PIT value is
the fraction of ensemble members below the realized observation. Under a
well-calibrated ensemble, PIT values should be approximately uniformly
distributed across simulation origins. We summarize deviation from uniformity by
the PIT score, defined as the Cram\'er-von Mises statistic of the empirical
PIT distribution against the uniform law; smaller values indicate better
calibration \citep{dawid1984prequential,diebold1998density,wilks2011statmethods}.
In the main text, we emphasize 90\% empirical coverage.

\subsubsection{Energy score}

For realized field \(y\) and ensemble members
\(x^{(1)},\ldots,x^{(M)}\), the energy score is
\[
\mathrm{ES}(F,y)
=
\frac{1}{M}\sum_{m=1}^{M}\|x^{(m)}-y\|_2
-
\frac{1}{2M^2}\sum_{m=1}^{M}\sum_{m'=1}^{M}
\|x^{(m)}-x^{(m')}\|_2.
\]
Lower values are better. In the reported experiments, ES is computed on the
same fixed fold-specific subset of 512 grid cells used for the variogram score.

\subsubsection{Variogram score}

For order \(\beta=0.5\), the variogram score is
\[
\mathrm{VS}(F,y)
=
\frac{1}{|\mathcal P|}
\sum_{(i,j)\in\mathcal P}
\left(
|y_i-y_j|^\beta
-
\frac{1}{M}\sum_{m=1}^{M}|x_i^{(m)}-x_j^{(m)}|^\beta
\right)^2,
\]
where \(\mathcal P=\{(i,j):1\le i<j\le d_{\mathrm{sub}}\}\) is the set of all
unordered pairs in the fixed \(d_{\mathrm{sub}}=512\) grid-cell subset. Thus
the weights are equal and normalized by the number of pairs. This score
emphasizes spatial dependence realism while keeping computation tractable.

\subsubsection{Warm-exceedance area CRPS}

Because warm extremes are of substantive interest, we report a CRPS diagnostic
for the spatial extent of warm-tail exceedance. For each meteorological season
\(s\) and grid cell \(i\), let \(u_{s,i}\) denote the training 95th-percentile
threshold of the reconstructed rank-\(K^*\) field. At origin \(o\) and lead
\(h\), define the realized warm-exceedance area fraction
\[
A_{o,h}
=
D^{-1}\sum_{i=1}^{D}
\mathbf{1}\{Y^{(K^*)}_{o+h,i}>u_{s(o+h),i}\}.
\]
For ensemble member \(m\), define \(A^{(m)}_{o,h}\) analogously. The
origin-specific ensemble CRPS of the scalar exceedance-area fraction is
\[
\mathrm{W95}_{o,h}
=
\frac{1}{M}\sum_{m=1}^{M}\left|A^{(m)}_{o,h}-A_{o,h}\right|
-
\frac{1}{2M^2}\sum_{m=1}^{M}\sum_{m'=1}^{M}
\left|A^{(m)}_{o,h}-A^{(m')}_{o,h}\right|.
\]
The reported W95 value at each lead is the average of
\(\mathrm{W95}_{o,h}\) over origins, folds, and regions. Lower values indicate
better calibration of the spatial extent of warm-tail
exceedance \citep{allen2024twcrps,gneiting2007scoring}.

\subsubsection{Persistence ratio}
Because the framework is intended for iterated simulation, we also report a
diagnostic relative to the naive persistence forecast:
\begin{equation}
\mathrm{Persist}(o,h)
=
\frac{\|\bar{Y}_{o+h}-Y_{o+h}^{(K^*)}\|_2^2}
{\|Y_{o}^{(K^*)}-Y_{o+h}^{(K^*)}\|_2^2},
\label{eq:persist_ratio}
\end{equation}
where $\bar{Y}_{o+h}=M^{-1}\sum_{m=1}^{M} Y_{o+h}^{(K^*)\,(m)}$ is the ensemble
mean. Values below one indicate improvement over persistence. We compute this
diagnostic at the origin level and then average over origins. Consequently, the
mean ratio can be sensitive to origins for which the persistence denominator is
very small; we therefore interpret it as a secondary dependence diagnostic,
complementing the energy and variogram scores rather than replacing them.

\subsubsection{Coefficient-space dependence reproduction}
Because the model is fitted in EOF coefficient space, we also compare observed
and simulated autocorrelation of the retained coefficients. These diagnostics
clarify whether field-level performance is accompanied by credible propagation
of the latent state itself.

In the main text, we summarize performance over short (<15 days), medium
(30-60 days), and longer (90-120 days) horizon bands. This keeps the
interpretation aligned with the main statistical question: how the balance
between dynamical propagation and climatological regularization changes with
lead time.

\subsection{Training residual diagnostics}
\label{sec:eval:resid}

These residual diagnostics are used to interpret, rather than rank, the
multi-step simulation results. In particular, they help distinguish whether
differences among \(\mathrm{M}_1\), \(\mathrm{M}_2\), and M3
arise from one-step innovation calibration, iterated propagation, or the
horizon-adaptive hybridization step. More granular checks, including whiteness
tests and coefficient-space calibration summaries, are reported in Appendix.

\begin{figure}[t]
\centering
\includegraphics[width=\textwidth]{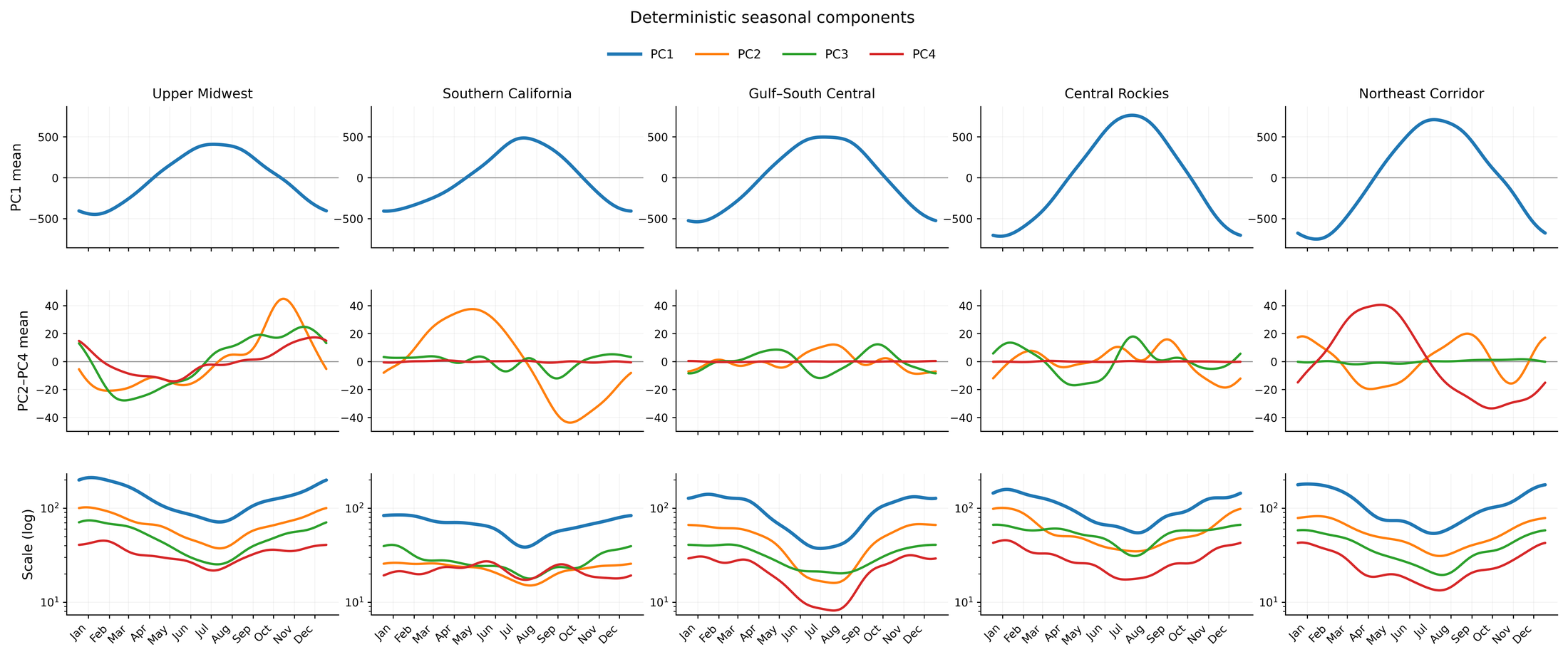}
\caption{Deterministic coefficient structure for the retained EOF
coefficients. Row 1 shows the fitted seasonal mean of PC 1, the dominant
domain-scale thermal mode. Row 2 shows the fitted seasonal means of PCs\,2--4,
which encode lower-variance regional contrast modes. Row 3 shows the
corresponding fitted seasonal scale functions on a logarithmic axis. Although
PC 1 has the largest absolute scale, PCs\,2--4 also exhibit substantial
relative seasonal scale modulation, supporting coefficient-specific seasonal
scale modeling rather than a common scalar seasonal adjustment
\citep{wallace2006atmo,wilks2011statmethods}.}
\label{fig:season_curves}
\end{figure}

\section{Results}
\label{sec:results}

Sections~\ref{sec:method} and~\ref{sec:eval} define the modeling and
verification framework. We collect the corresponding empirical checks and
out-of-sample verification results here so that structural diagnostics,
simulation scores, and horizon-dependent trade-offs can be interpreted
together. The results are organized around two linked goals. The inferential
goal is to learn whether the retained EOF coefficients form a stable and
interpretable temperature state, and to estimate the seasonal, persistence,
dispersion, tail, low-frequency, and hybrid-shrinkage structure of that state.
The predictive goal is to test whether this estimated state can be propagated
under rolling origins to generate calibrated and spatially coherent temperature
fields. We report results for Upper Midwest, Southern California,
Gulf--South Central, Central Rockies, and Northeast Corridor under the strict
rolling-origin design described in Sect.~\ref{sec:exp} and detailed in
Appendix~\ref{app:implementation}. Unless otherwise stated, verification
metrics are computed for the reconstructed rank-$K^*$ field $Y^{(K^*)}$ and
then aggregated across the four folds.

The main text compares seven models: the climatology benchmark
\textbf{CLIM}; three dynamical simulators $\mathrm{M}_1$, $\mathrm{M}_2$,
$\mathrm{M}_3$; and their horizon-adaptive hybrids
\textbf{Hybrid-$\mathrm{M}_1$}, \textbf{Hybrid-$\mathrm{M}_2$},
\textbf{Hybrid-$\mathrm{M}_3$}. These represent a progression from a
resampling-based benchmark to increasingly structured dynamical simulations of
the retained low-rank state: $\mathrm{M}_1$ uses a constant innovation
covariance, $\mathrm{M}_2$ introduces tuned EWMA-based marginal volatility, and
$\mathrm{M}_3$ uses an EWMA-regularized GRU volatility correction. Each raw
dynamical model is paired with a lead-adaptive hybrid that regularizes the
ensemble toward climatology as horizon increases.

\subsection{The retained EOF state is compact, coherent, and stable}

A first requirement of the framework is that the retained EOF state represent a
genuinely low-dimensional and statistically stable component of the regional
temperature field. Table~\ref{tab:m1_structure} shows that this condition is
met in all five domains. The reported experiments use \(K^*=4\), for which the
cumulative explained variance exceeds 95\% in every region
(96.7-97.5\%), while EOF 1 alone explains 89--91\% of the variance. Thus
the retained state is compact but not one-dimensional: PC 1 captures the
dominant domain-scale thermal mode, while PCs\,2--4 retain lower-variance
regional contrast modes used by the dynamical simulator. This choice keeps the
state dimension small enough for stable multivariate dynamics while preserving
the fourth retained EOF used throughout the simulation and verification
pipeline.

The fitted seasonal VARX lag orders range from $p=2$ (Southern California) to $p=4$
(Central Rockies, Gulf--South Central). All fitted companion matrices have spectral radii well
below one, confirming dynamic stability of the iterated state propagation. \begin{table}[t]
\centering
\caption{Regional EOF structure and fitted model parameters. \(K^*=4\) is the
retained rank used in the reported experiments; \(\mathrm{CEV}_4\) is the
cumulative explained variance of the first four EOFs. \(p\) is the selected
seasonal VAR lag order; \(\rho_{\max}\) is the maximum spectral radius of
the fitted seasonal VARX companion matrix; and \(\bar{\nu}\) is the
fold-averaged median fitted Student-\(t\) degrees of freedom across retained
coefficients.}
\label{tab:m1_structure}
\small
\begin{tabular}{lrrrrrr}
\toprule
Region & $K^*$ & EV$_1$\,(\%) & CEV$_4$\,(\%) & $p$ & $\rho_{\max}$ & $\bar{\nu}$ \\
\midrule
Upper Midwest & 4 & 90.1 & 97.1 & 3 & 0.72 & 15.8 \\
Central Rockies & 4 & 89.3 & 97.1 & 4 & 0.79 & 18.5 \\
Gulf--South Central & 4 & 90.0 & 97.1 & 4 & 0.76 & 11.2 \\
Southern California & 4 & 89.4 & 96.7 & 2 & 0.88 & 16.1 \\
Northeast Corridor & 4 & 90.5 & 97.5 & 3 & 0.76 & 17.3 \\
\bottomrule
\end{tabular}
\end{table}

Figure~\ref{fig:eof_maps} shows that the leading modes are physically coherent.
Because EOF signs are arbitrary, interpretation rests on spatial contrasts
rather than on sign alone. EOF 1 acts as a spatially weighted domain-scale
thermal mode in every region: all loadings have the same sign, so the
coefficient amplitude describes a broadly coherent warm-versus-cold fluctuation
relative to the training mean. EOF\,2 typically introduces a large-scale
dipole, often north-south or coast-interior, while EOF\,3 captures more
region-specific secondary structure. In the coastal domains, especially
Southern California and Northeast Corridor, higher-order loadings reflect sharp ocean-land
transitions, supporting the interpretation that the retained EOFs are stable
large-scale modes with clear climatological meaning.

Table~\ref{tab:m2_acf} shows that the retained coefficients also have distinct
temporal persistence. PC 1 is the most persistent mode in every region, while
higher-order modes exhibit region-specific ordering. Thus the reduced-rank
state is not only compact, but dynamically heterogeneous in a way that must be
captured by the dependence model. Innovation tails vary meaningfully across
regions: Gulf--South Central shows the heaviest tails (PC 1 df $= 7.6$), while other
regions range from 12 to 33 (Appendix Table~\ref{tab:m3_tdf}), supporting the
use of a flexible Student-t innovation distribution.

\begin{table}[t]
\centering
\caption{Lag-1 autocorrelation of standardized EOF coefficients by region.
PC 1 is consistently the most persistent; higher-order modes show
region-specific ordering.}
\label{tab:m2_acf}
\small
\begin{tabular}{lrrrr}
\toprule
Region & PC 1 & PC\,2 & PC\,3 & PC\,4 \\
\midrule
Upper Midwest & 0.750 & 0.572 & 0.418 & 0.275 \\
Central Rockies & 0.737 & 0.599 & 0.530 & 0.524 \\
Gulf--South Central & 0.749 & 0.491 & 0.631 & 0.393 \\
Southern California & 0.798 & 0.707 & 0.510 & 0.572 \\
Northeast Corridor & 0.677 & 0.576 & 0.359 & 0.298 \\
\bottomrule
\end{tabular}
\end{table}

\begin{table}[t]
\centering
\caption{Estimated reduced-state properties by region not already reported in
Table~\ref{tab:m2_acf}. Drift counts give the number of rolling folds in which
the screened PC\,1 drift term was selected. Scale ratio is the annual
max-to-min ratio of the fitted PC\,1 seasonal scale. \(\bar{\nu}\) is the
fold-averaged median fitted Student-\(t\) degrees of freedom across retained
coefficients. \(\bar{\tau}_{\mathrm{M3}}\) is the mean fitted hybrid shrinkage time
scale for Hybrid-\(\mathrm{M}_3\).}
\label{tab:estimated_state_properties}
\small
\setlength{\tabcolsep}{5pt}
\begin{tabular}{lrrrr}
\toprule
Region & PC~1 drift & PC~1 scale ratio & \(\bar{\nu}\) &
\(\bar{\tau}_{\mathrm{M3}}\) \\
\midrule
Upper Midwest & 3/4 & 2.97 & 15.8 & 217 \\
Central Rockies & 4/4 & 2.92 & 18.5 & 173 \\
Gulf--South Central & 3/4 & 3.79 & 11.2 & 262 \\
Southern California & 4/4 & 2.22 & 16.1 & 310 \\
Northeast Corridor & 2/4 & 3.35 & 17.3 & 180 \\
\bottomrule
\end{tabular}
\end{table}

Table~\ref{tab:estimated_state_properties} summarizes drift selection,
seasonal scale amplitude, innovation-tail heaviness, and hybrid shrinkage. The
fitted scale ratios show substantial annual modulation of the dominant mode.
PC\,1 drift is selected in most folds and is positive whenever selected,
consistent with the leading coefficient representing a warming-sensitive
regional thermal amplitude. The fitted Student-\(t\) degrees of freedom are
smallest in Gulf--South Central, indicating heavier innovation tails there,
while the hybrid shrinkage time scale \(\tau\) varies strongly by region,
showing that the transition rate from the parametric generator toward
state-matched climatological continuation is domain dependent rather than
universal. Together with the persistence results in Table~\ref{tab:m2_acf},
these parameter estimates provide interpretable regional summaries in addition
to producing forecast ensembles.

Finally, Fig.~\ref{fig:stab_max_angle} shows that the retained EOF subspaces
remain reasonably stable under rolling refits, although the degree of stability
varies by region. Stability is strongest in Northeast Corridor and Central Rockies, while
Upper Midwest has the largest cross-window variability, with a median maximum
principal angle near 6 degrees and some refits exceeding 9 degrees.
Southern California and Gulf--South Central are intermediate. Even in the least stable case, the
angles remain modest relative to a complete rotation of the retained subspace.

\subsection{The retained coefficients have distinct climatological structure}

Figure~\ref{fig:season_curves} shows that all four retained coefficients have
clear deterministic structure in mean and scale. PC 1 exhibits the dominant
annual mean cycle in every region, consistent with its interpretation as a
domain-scale thermal mode. PCs\,2--4 have smaller mean amplitudes but are not
residual noise: they encode region-specific contrast modes whose seasonal
structure differs by domain.

The fitted scale functions are equally important. PC 1 has the largest
absolute scale, but the log-scale display shows that PCs\,2--4 can undergo
seasonal scale modulation of comparable relative amplitude. Across regions,
annual max-to-min scale ratios are roughly 2--4 for several retained
coefficients. This supports the coefficient-specific decomposition in
Eqs.~\eqref{eq:mk_final}--\eqref{eq:sk_final}: a common scalar seasonal
variance correction would miss important heterogeneity across the retained
state.

Taken together, Figs.~\ref{fig:eof_maps} and~\ref{fig:season_curves} show
that the retained EOF state has a domain-specific climatology. The loading maps
are spatially coherent, and the coefficient amplitudes vary through the annual
cycle in systematic but region-dependent ways. The estimated deterministic
parameters therefore provide more than a normalization step: they show how the
dominant regional thermal mode and the lower-variance contrast modes change in
mean and volatility over the seasonal cycle.

Table~\ref{tab:covariate_assoc_main} lists the low-frequency associations
that pass the fold-stability screen. Its purpose is to show where the covariate
block contributes reproducible, though modest, secondary structure after the
seasonal harmonics and screened drift have been removed.

\begin{table}[t]
\centering
\caption{Stable low-frequency covariate associations in retained EOF
coefficient space. The median effect is the fitted change in EOF-coefficient amplitude for a
one-standard-deviation increase in the training-standardized covariate; the
coefficient amplitude remains on the scale of the EOF projection. Effects are estimated within season after
removing harmonic seasonal structure and the screened drift term from the
corresponding coefficient. CI rate is the fraction of folds for which the
bootstrap interval excludes zero. \(R^2_{\mathrm{cov}}\) is the mean
within-season explanatory fraction of the covariate block.}
\label{tab:covariate_assoc_main}
\small
\setlength{\tabcolsep}{4.5pt}
\begin{tabular}{lllrrrr}
\toprule
Region & PC & Season & Covariate & Median effect & CI rate & \(R^2_{\mathrm{cov}}\) \\
\midrule
Gulf--South Central & PC1 & JJA & CO$_2$ &  69.3 & 0.75 & 0.081 \\
Gulf--South Central & PC1 & JJA & AMO    &  11.5 & 1.00 & 0.081 \\
Upper Midwest & PC1 & JJA & ONI    & -26.8 & 1.00 & 0.065 \\
Upper Midwest & PC1 & JJA & AMO    &  19.2 & 1.00 & 0.065 \\
Southern California & PC2 & SON & ONI    &  -9.6 & 1.00 & 0.063 \\
Northeast Corridor & PC1 & JJA & ONI    & -24.2 & 1.00 & 0.058 \\
Northeast Corridor & PC1 & JJA & AMO    &  12.6 & 1.00 & 0.058 \\
Central Rockies & PC1 & JJA & ONI    & -13.1 & 1.00 & 0.051 \\
Southern California & PC2 & JJA & ONI    &  -6.9 & 1.00 & 0.045 \\
Southern California & PC2 & DJF & CO$_2$ & -43.5 & 0.75 & 0.032 \\
Southern California & PC2 & DJF & ONI    &  -7.7 & 1.00 & 0.032 \\
Central Rockies & PC3 & DJF & CO$_2$ & 163.5 & 1.00 & 0.023 \\
Central Rockies & PC3 & DJF & AMO    & -21.4 & 1.00 & 0.023 \\
\bottomrule
\end{tabular}
\end{table}

The low-frequency covariate associations are modest compared with the
deterministic seasonal structure. The largest stable \(R^2_{\mathrm{cov}}\)
values are below 0.1, so CO$_2$, ONI, and AMO should be viewed as secondary
adjustments rather than primary drivers of coefficient variability. The
selected associations are concentrated in JJA and DJF: no MAM association and
only one SON association passes the stated stability screen. This asymmetry is
a feature of the screened summary, not evidence that spring and autumn effects
are absent. The estimates are useful inferentially because they identify where
low-frequency climate indices enter the retained state, but because the indices
and the screened drift adjustment are correlated over the historical record,
the signs of individual low-frequency coefficients should not be interpreted
causally.

\subsection{Residual diagnostics support a structured innovation specification}

Figure~\ref{fig:acf_train} shows that the seasonal VARX layer removes a
substantial fraction of the serial dependence present in the standardized
coefficient process, but it does not render the residual process featureless.
Remaining scale variation and region-specific departures from ideal whiteness
motivate the innovation specifications used in $\mathrm{M}_2$ and
$\mathrm{M}_3$: time-varying marginal scale, seasonal cross-mode dependence,
and dispersion calibration. Appendix Table~\ref{tab:s3_whitening} shows that
the seasonal correlation factor removes most off-diagonal cross-mode
dependence, and that $\mathrm{M}_2$ has the lowest ARCH rejection rate.
However, improved one-step diagnostics do not automatically imply better
multi-step field simulation. This mismatch is itself informative: residual
whitening, marginal calibration, and iterated field generation are related but
not identical targets, so rolling-origin verification is needed to evaluate the
predictive simulator. The implications for iterated propagation are discussed
in Sect.~\ref{sec:discussion}.

\subsection{Simulation quality is strongly horizon dependent}

\begin{table}[t]
\centering
\caption{Horizon-dependent verification for the main comparison set. Energy
score is lower-is-better and Cov$_{90}$ is empirical 90\% coverage. Values are
averaged across five regions, four rolling folds, and simulation origins.}
\label{tab:m5_horizon}
\small
\setlength{\tabcolsep}{3.0pt}
\begin{tabular}{llrrrrrrrr}
\toprule
Model & Metric & $h{=}1$ & $h{=}3$ & $h{=}7$ & $h{=}14$ & $h{=}30$ & $h{=}60$ & $h{=}90$ & $h{=}120$ \\
\midrule
CLIM & Energy & 52.9 & 53.1 & 54.3 & 54.8 & 52.8 & 51.5 & 53.2 & 53.9 \\
CLIM & Cov$_{90}$ & 0.881 & 0.875 & 0.867 & 0.860 & 0.876 & 0.870 & 0.874 & 0.862 \\
\addlinespace
$\mathrm{M}_1$ & Energy & 33.5 & 50.3 & 52.8 & 53.4 & 52.0 & 51.3 & 52.4 & 53.3 \\
$\mathrm{M}_1$ & Cov$_{90}$ & 0.932 & 0.923 & 0.929 & 0.921 & 0.929 & 0.924 & 0.933 & 0.921 \\
\addlinespace
$\mathrm{M}_3$ & Energy & 33.4 & 50.3 & 52.5 & 52.5 & 51.8 & 50.9 & 52.1 & 53.0 \\
$\mathrm{M}_3$ & Cov$_{90}$ & 0.909 & 0.902 & 0.902 & 0.896 & 0.902 & 0.903 & 0.911 & 0.899 \\
\addlinespace
Hyb-$\mathrm{M}_3$ & Energy & 33.4 & 50.3 & 52.5 & 52.4 & 51.3 & 50.4 & 51.4 & 51.6 \\
Hyb-$\mathrm{M}_3$ & Cov$_{90}$ & 0.909 & 0.902 & 0.902 & 0.892 & 0.896 & 0.887 & 0.881 & 0.861 \\
\bottomrule
\end{tabular}
\end{table}

Tables~\ref{tab:m5_horizon} and~\ref{tab:s9_band_summary} show clear horizon
dependence. The one-step forecasts give the largest field-score gain:
energy score decreases from 52.9 for CLIM to about 33.4 for the dynamical
models. This behavior is consistent with the intended rolling-origin convention,
where the observed state at the origin is used to simulate the first future day
\(t_0+1\). Because the hybrid branch is not allowed before lead 7, each hybrid
is identical to its parent dynamical model at horizons 1, 3, and 7.

Over 3--14~days, \(\mathrm{M}_3\) and Hybrid-\(\mathrm{M}_3\) provide the
strongest field-score performance while remaining close to nominal 90\%
coverage. \(\mathrm{M}_1\) and \(\mathrm{M}_2\) are more conservative, with
coverage around 0.92--0.93 in this band. At medium horizons (30--60~days),
hybridization becomes clearly beneficial: Hybrid-\(\mathrm{M}_2\) gives the
lowest band-averaged energy score (50.7), with the other hybrids close behind.
At 90--120~days, all three hybrids outperform CLIM and their corresponding raw
dynamical models in energy score; Hybrid-\(\mathrm{M}_1\), Hybrid-\(\mathrm{M}_2\),
and Hybrid-\(\mathrm{M}_3\) are close, with energy scores between 51.43 and
51.49. The incremental hybridization gains relative to the corresponding parent
models are modest in absolute terms (generally about 1--3\% across the medium-
and long-horizon bands), although they are consistent across all three bases;
the much larger practical gain occurs at one-day lead through conditioning on
the observed origin state.

The W95 diagnostic gives a different view from the multivariate field scores.
At one-day lead, the dynamical forecasts improve W95 relative to
CLIM, reflecting the value of conditioning on the observed origin state. At
longer leads, CLIM remains a strong marginal-tail benchmark because it directly
resamples historical warm-tail realizations. This is expected: the retained
rank-\(K^*=4\) field is designed to represent dominant large-scale structure,
whereas localized residual variability outside the retained subspace can matter
for exceedance-area diagnostics during heat events. We therefore report W95 as
a calibration trade-off rather than as the primary criterion for the proposed
state-space simulator, whose main target is coherent conditional simulation of
the reconstructed rank-\(K^*\) regional field. The persistence ratio further
distinguishes the models but should be interpreted cautiously because it is an
average of origin-level ratios with a sometimes small persistence denominator:
at 90--120~days, all hybrids achieve mean ratios below 1.0, with
Hybrid-\(\mathrm{M}_3\) lowest at 0.78.

\begin{table}[t]
\centering
\caption{Horizon-band summary for all seven models. Energy score and variogram
score (VS) are lower-is-better multivariate field scores; coverage is empirical
90\% coverage; W95 is the CRPS of the warm-exceedance area fraction; and
Persist is the persistence ratio. Values are averaged across regions, folds,
and simulation origins.}
\label{tab:s9_band_summary}
\small
\setlength{\tabcolsep}{4.0pt}
\begin{tabular}{llrrrrr}
\toprule
Band & Model & Energy & VS & Cov$_{90}$ & W95 & Persist \\
\midrule
1\,d & CLIM & 52.9 & 0.366 & 0.881 & 0.074 & 10.68 \\
1\,d & $\mathrm{M}_1$ & 33.5 & 0.248 & 0.932 & 0.054 & 3.61 \\
1\,d & $\mathrm{M}_2$ & 33.4 & 0.244 & 0.931 & 0.053 & 3.38 \\
1\,d & $\mathrm{M}_3$ & 33.4 & 0.247 & 0.909 & 0.051 & 3.13 \\
1\,d & Hyb-$\mathrm{M}_1$ & 33.5 & 0.248 & 0.932 & 0.054 & 3.61 \\
1\,d & Hyb-$\mathrm{M}_2$ & 33.4 & 0.244 & 0.931 & 0.053 & 3.38 \\
1\,d & Hyb-$\mathrm{M}_3$ & 33.4 & 0.247 & 0.909 & 0.051 & 3.13 \\
\addlinespace
3--14\,d & CLIM & 54.1 & 0.369 & 0.867 & 0.091 & 4.28 \\
3--14\,d & $\mathrm{M}_1$ & 52.2 & 0.368 & 0.925 & 0.105 & 4.20 \\
3--14\,d & $\mathrm{M}_2$ & 52.4 & 0.369 & 0.931 & 0.109 & 4.35 \\
3--14\,d & $\mathrm{M}_3$ & 51.8 & 0.367 & 0.900 & 0.099 & 3.67 \\
3--14\,d & Hyb-$\mathrm{M}_1$ & 52.1 & 0.367 & 0.923 & 0.105 & 4.16 \\
3--14\,d & Hyb-$\mathrm{M}_2$ & 52.4 & 0.368 & 0.929 & 0.108 & 4.32 \\
3--14\,d & Hyb-$\mathrm{M}_3$ & 51.7 & 0.366 & 0.899 & 0.099 & 3.65 \\
\addlinespace
30--60\,d & CLIM & 52.1 & 0.358 & 0.873 & 0.082 & 1.46 \\
30--60\,d & $\mathrm{M}_1$ & 51.7 & 0.362 & 0.927 & 0.106 & 1.60 \\
30--60\,d & $\mathrm{M}_2$ & 51.7 & 0.365 & 0.928 & 0.109 & 1.72 \\
30--60\,d & $\mathrm{M}_3$ & 51.3 & 0.362 & 0.902 & 0.100 & 1.43 \\
30--60\,d & Hyb-$\mathrm{M}_1$ & 50.8 & 0.357 & 0.908 & 0.097 & 1.47 \\
30--60\,d & Hyb-$\mathrm{M}_2$ & 50.7 & 0.358 & 0.910 & 0.099 & 1.53 \\
30--60\,d & Hyb-$\mathrm{M}_3$ & 50.8 & 0.358 & 0.891 & 0.094 & 1.37 \\
\addlinespace
90--120\,d & CLIM & 53.6 & 0.368 & 0.868 & 0.096 & 0.90 \\
90--120\,d & $\mathrm{M}_1$ & 52.8 & 0.374 & 0.927 & 0.116 & 0.98 \\
90--120\,d & $\mathrm{M}_2$ & 53.0 & 0.372 & 0.919 & 0.118 & 1.12 \\
90--120\,d & $\mathrm{M}_3$ & 52.6 & 0.372 & 0.905 & 0.110 & 0.92 \\
90--120\,d & Hyb-$\mathrm{M}_1$ & 51.4 & 0.363 & 0.875 & 0.100 & 0.79 \\
90--120\,d & Hyb-$\mathrm{M}_2$ & 51.5 & 0.362 & 0.875 & 0.101 & 0.81 \\
90--120\,d & Hyb-$\mathrm{M}_3$ & 51.5 & 0.361 & 0.871 & 0.100 & 0.78 \\
\bottomrule
\end{tabular}
\end{table}

\subsection{Dynamical models improve short-lead state persistence}

\begin{table}[t]
\centering
\caption{Mean member-level coefficient-space autocorrelation: observed versus
simulated. Dynamical models preserve substantially more trajectory-level
temporal dependence than CLIM at short horizons. By \(h=60\), differences
across simulators narrow.}
\label{tab:m6_pc_acf}
\small
\begin{tabular}{lrrrrrrr}
\toprule
Model & $h{=}1$ & $h{=}3$ & $h{=}7$ & $h{=}14$ & $h{=}30$ & $h{=}60$ & $h{=}90$ \\
\midrule
Observed & 0.643 & 0.429 & 0.363 & 0.324 & 0.272 & 0.191 & $-$0.001 \\
\midrule
CLIM & 0.321 & 0.323 & 0.319 & 0.311 & 0.268 & 0.145 & $-$0.002 \\
$\mathrm{M}_1$ & 0.623 & 0.346 & 0.309 & 0.298 & 0.253 & 0.133 & $-$0.009 \\
$\mathrm{M}_3$ & 0.615 & 0.346 & 0.317 & 0.298 & 0.257 & 0.134 & $-$0.006 \\
Hyb-$\mathrm{M}_3$ & 0.615 & 0.346 & 0.317 & 0.300 & 0.263 & 0.136 & $-$0.004 \\
\bottomrule
\end{tabular}
\end{table}

The values in Table~\ref{tab:m6_pc_acf} are from the final configuration
documented in Appendix~\ref{app:implementation}, including the
coefficient-specific persistence vector \(\boldsymbol{\rho}=(0.1,0,0,0)\),
the descending lag cap, and \(M=100\) ensemble members. Earlier development
runs used different regularization settings and are not pooled with these
reported values.

Table~\ref{tab:m6_pc_acf} shows that the dynamical models reproduce the
trajectory-level temporal dependence of the retained state more faithfully than
CLIM at very short leads. At \(h=1\), the dynamical models recover nearly all
of the observed mean coefficient-space autocorrelation (about 0.62 versus 0.64
observed), whereas CLIM captures only about half. The improvement is clearest
for PC 1: the per-coefficient breakdown in Appendix
Table~\ref{tab:m7_perpc_acf} shows that the dynamical models reproduce PC 1
autocorrelation very closely at \(h=1\). For higher-order coefficients and
longer leads, the differences narrow, reflecting the fact that regularized
linear VARX propagation cannot fully reproduce all retained-mode persistence
scales simultaneously.

The coefficient-space autocorrelation also clarifies the role of the hybrid.
Because branching is not permitted before lead 7, Hybrid-M3 and
M3 are identical at the first three reported leads. At longer
leads, the hybrid gradually moves part of the ensemble to state-conditioned
climatological continuations. This stabilizes field-level scores but slightly
damps medium-range coefficient persistence, especially by \(h=60\), where the
hybrid has lower member-level autocorrelation than the observed retained state.

\subsection{Hybridization is the dominant source of improvement at medium and long horizons}

\begin{table}[t]
\centering
\caption{Hybridization gains by horizon band. Negative values indicate
improvement for the lower-is-better scores. The trajectory-branching hybrid
improves energy score and variogram score for every base model at medium and
long horizons.}
\label{tab:m8_hybrid_gains}
\small
\setlength{\tabcolsep}{5pt}
\begin{tabular}{lrrrr}
\toprule
Pair & \multicolumn{2}{c}{Mid (30--60\,d)} & \multicolumn{2}{c}{Long (90--120\,d)} \\
\cmidrule(lr){2-3} \cmidrule(lr){4-5}
 & $\Delta$ES & $\Delta$VS & $\Delta$ES & $\Delta$VS \\
\midrule
$\mathrm{M}_1\!\to\!$Hyb-$\mathrm{M}_1$ & $-$0.81 & $-$0.005 & $-$1.42 & $-$0.011 \\
$\mathrm{M}_2\!\to\!$Hyb-$\mathrm{M}_2$ & $-$0.92 & $-$0.006 & $-$1.49 & $-$0.010 \\
$\mathrm{M}_3\!\to\!$Hyb-$\mathrm{M}_3$ & $-$0.49 & $-$0.004 & $-$1.08 & $-$0.011 \\
\bottomrule
\end{tabular}
\end{table}

Table~\ref{tab:m8_hybrid_gains} shows that hybridization improves energy score
and variogram score for every base model at medium and long horizons, while
leaving short-range performance unchanged by construction. The largest
long-horizon energy gain occurs for \(\mathrm{M}_2\)
(\(\Delta\mathrm{ES}=-1.49\) at 90--120~days), but all three bases benefit.
At 90--120~days, the three hybrids have very similar energy scores
(51.43--51.49), so the main empirical finding is not a decisive win by one
volatility specification. Instead, the robust result is that state-matched
trajectory branching provides medium- and long-horizon regularization that raw
dynamical propagation lacks. Hybrid-\(\mathrm{M}_3\) retains the best
long-horizon persistence ratio (0.78), whereas Hybrid-\(\mathrm{M}_1\) and
Hybrid-\(\mathrm{M}_2\) are slightly stronger in some energy-score averages and
remain close to nominal coverage.

To assess the robustness of these score differences, we also compute paired
origin-block bootstrap intervals by resampling forecast origins within each
region--fold block and recomputing the band-averaged paired differences
(Table~\ref{tab:bootstrap_hybrid}). The intervals should be interpreted as
descriptive dependence-aware uncertainty summaries, not as independent-fold
standard errors, because the rolling test periods overlap. They nevertheless
show that the medium- and long-horizon hybrid gains in energy score and
variogram score are consistently separated from zero.

\begin{table}[t]
\centering
\caption{Paired origin-block bootstrap intervals for hybridization gains. Each
entry is the hybrid score minus the corresponding parent dynamical-model score;
negative values indicate improvement. Intervals are 95\% percentile intervals
from resampling forecast origins within each region--fold block.}
\label{tab:bootstrap_hybrid}
\small
\setlength{\tabcolsep}{4pt}
\resizebox{\textwidth}{!}{%
\begin{tabular}{lrrrr}
\toprule
Pair & \multicolumn{2}{c}{Mid (30--60\,d)} & \multicolumn{2}{c}{Long (90--120\,d)} \\
\cmidrule(lr){2-3} \cmidrule(lr){4-5}
 & $\Delta$ES [95\% int.] & $\Delta$VS [95\% int.] & $\Delta$ES [95\% int.] & $\Delta$VS [95\% int.] \\
\midrule
$\mathrm{M}_1\!\to\!$Hyb-$\mathrm{M}_1$ & $-0.81$ [$-1.04,-0.57$] & $-0.0049$ [$-0.0070,-0.0025$] & $-1.42$ [$-1.82,-0.97$] & $-0.0111$ [$-0.0151,-0.0072$] \\
$\mathrm{M}_2\!\to\!$Hyb-$\mathrm{M}_2$ & $-0.92$ [$-1.15,-0.68$] & $-0.0064$ [$-0.0087,-0.0041$] & $-1.49$ [$-1.92,-1.05$] & $-0.0103$ [$-0.0139,-0.0064$] \\
$\mathrm{M}_3\!\to\!$Hyb-$\mathrm{M}_3$ & $-0.49$ [$-0.74,-0.26$] & $-0.0045$ [$-0.0066,-0.0023$] & $-1.08$ [$-1.56,-0.62$] & $-0.0107$ [$-0.0149,-0.0068$] \\
\bottomrule
\end{tabular}%
}
\end{table}

The fitted hybrid shrinkage time scale \(\tau\) for \(\mathrm{M}_3\)
(Appendix Table~\ref{tab:m9_hybrid_tau}) varies substantially across regions,
from a mean of about 173 days in Central Rockies to about 310 days in Southern California.
These values should not be interpreted as physical predictability horizons of
several months. Rather, \(\tau\) parameterizes the rate at which the hybrid
switches from the parametric low-rank generator toward state-matched historical
continuations. Large values therefore indicate that the fitted seasonal--drift
state-space generator remains competitive with block resampling for longer
lead times in that region. At \(h=120\) days, the expected fraction of ensemble
members still following the dynamical generator is
\(\exp\{-(120-h_{\min})/\tau\}\). With \(h_{\min}=7\), this ranges from
roughly 52\% in Central Rockies to roughly 69\% in Southern California.
The fact that the hybrids outperform CLIM at \(h=120\) under the energy and
variogram scores indicates that a partially parametric, partially resampled
ensemble retains useful large-scale structure beyond unconditioned block
resampling alone.

\section{Discussion and conclusions}
\label{sec:discussion}

The results support three main conclusions for the reconstructed rank-\(K^*=4\)
regional temperature field. Together, these conclusions show that the framework
is both inferential and predictive: it estimates interpretable structure in a
low-dimensional temperature state and uses that estimated structure for
multi-horizon probabilistic field simulation.

First, the retained EOF representation functions as a substantive statistical
state rather than merely as a compression device. Across all five regions, the
first four EOFs explain about 96.7--97.5\% of centered-field variance, the
retained subspaces remain stable under rolling refits, and the loading maps
identify coherent large-scale modes with clear climatological interpretation.
Because these fractions are computed for training-centered rather than
deseasonalized anomaly fields, they should not be compared directly with
anomaly-field PCA fractions; the centered-field definition matches the
generator's projected-field target and retains the large-scale seasonal signal.
PC 1 is a domain-scale thermal mode, while PCs\,2--4 represent regional
contrast modes. The fitted seasonal mean, seasonal scale, screened drift,
persistence, and tail parameters show that the retained coefficients have
interpretable climatology and dynamics, not just numerical dimension-reduction
value.

Second, model performance is strongly horizon dependent. At one-day lead, the forecasts show large gains from the
dynamical state-space models: energy score decreases from 52.9 for CLIM to
about 33.4 for the dynamical models. At 3--14~days, M3 and
Hybrid-M3 give the lowest band-averaged energy scores while
remaining close to nominal 90\% coverage. This distinction is important:
conservative one-step innovation diagnostics do not by themselves guarantee the
best iterated field simulation.

Third, hybridization is the main source of improvement beyond the short range.
The hybrids are identical to their parent dynamical models through the minimum
branch lead, but at 30--60 and 90--120~days the trajectory-branching hybrids
improve energy and variogram scores for all base models. These gains come with
a calibration trade-off. Hybrid-M3 gives the strongest
persistence-ratio performance and competitive variogram scores, whereas
Hybrid-\(\mathrm{M}_1\) and Hybrid-\(\mathrm{M}_2\) are slightly stronger in
some energy-score averages and remain close to nominal coverage. CLIM remains a
useful marginal warm-tail benchmark because it resamples historical extremes,
but this does not make it preferable for the paper's primary target: coherent
conditional simulation of the reconstructed rank-\(K^*\) regional field.
A related limitation is that the CLIM reference samples from the historical
training archive without an explicit low-frequency drift adjustment. Its
slight undercoverage, and the tendency of hybrids to approach that coverage at
long leads, are therefore consistent with nonstationarity in the historical
record. A trend-adjusted climatological reference would be a useful additional
ablation in future work for separating the contributions of slow mean change,
state conditioning, and trajectory-level hybridization.

These results do not identify a single universally best simulator for every
verification criterion. Rather, they show that an interpretable reduced-rank
state-space framework can support a coherent division of labor: stable EOF
coefficients provide the retained regional state, regularized VARX dynamics
propagate short- to medium-range dependence, structured innovations control
residual dispersion, and trajectory-level climatological hybridization
stabilizes the predictive distribution once dynamical information weakens. The
estimated parameters also provide scientifically useful summaries of regional
heterogeneity: Gulf--South Central has the heaviest retained-state innovation
tails, Southern California retains the longest hybrid dynamical time scale, and
Central Rockies and Gulf--South Central require longer VARX memory. These are
not ancillary diagnostics; they are part of the inferential value of the
state-space formulation. All
claims concern the reconstructed rank-\(K^*=4\) field. Variability outside the
retained EOF subspace remains unmodeled, so the simulator should not be used
directly for local microclimate or site-scale extreme-risk estimates without
additional downscaling or a stochastic residual spatial component. Extending
the framework to include a structured higher-rank residual field, or to model
multiple weather variables jointly, would be natural next steps.


\appendix


\section{Mathematical details}
\label{app:math}

This appendix collects the mathematical specification of the model components.
The notation follows the main text.

\subsection{Low-rank state-space formulation}

Let $Y_t \in \mathbb{R}^D$ denote the daily temperature field, with $D=4096$.
Let $\Phi = [\phi_1,\ldots,\phi_{K^*}] \in \mathbb{R}^{D\times K^*}$ be the
retained EOF loading matrix. The retained coefficient vector and rank-$K^*$
reconstruction are
\begin{equation}
a_t = \Phi^\top (Y_t - \mu_{\mathrm{tr}}),
\qquad
Y_t^{(K^*)} = \mu_{\mathrm{tr}} + \Phi\, a_t.
\end{equation}
The standardized coefficient state is
\begin{equation}
\varepsilon_{t,k}
= \frac{a_{t,k}-m_k(d_t)}{s_k(d_t)},
\qquad
\varepsilon_t = (\varepsilon_{t,1},\ldots,\varepsilon_{t,K^*})^\top,
\end{equation}
so that $a_t = m(d_t) + s(d_t) \odot \varepsilon_t$.

\subsection{Seasonal VARX propagation with structured regularization}

The standardized state is propagated by a seasonal VARX($p$),
\begin{equation}
\varepsilon_t
= c_{q(t)} + \sum_{i=1}^p A_{i,q(t)}\,\varepsilon_{t-i}
  + B_{q(t)}\,\widetilde{x}_t + u_t,
\end{equation}
where $q(t)\in\{\mathrm{DJF},\mathrm{MAM},\mathrm{JJA},\mathrm{SON}\}$
denotes the meteorological season. The fitted one-step residual is
$\widehat{u}_t = \varepsilon_t - \widehat{c}_{q(t)}
- \sum_i \widehat{A}_{i,q(t)}\varepsilon_{t-i}
- \widehat{B}_{q(t)}\widetilde{x}_t$.

Estimation uses a structured ridge penalty. Let
$\sigma_k^2 = \mathrm{Var}(\varepsilon_{t,k})$ denote the marginal variance of
the $k$-th standardized coefficient. The ridge target for the $(j,k)$ entry of
the autoregressive matrices is proportional to $1/\sigma_k^2$, so that
lower-variance higher-order modes receive stronger shrinkage. The effective
penalty for coefficient $k$ is $\lambda_{\mathrm{target}}/\sigma_k^2$, where
$\lambda_{\mathrm{target}}$ is a global strength parameter.

A descending maximum-lag rule is used: mode \(k\) has maximum lag
\(p_k=\max\{p-k+1,1\}\). Thus the leading mode uses the full fitted lag
\(p\), each successive mode uses one fewer lag, and coefficients beyond the
mode-specific maximum are set to zero.

\subsection{Innovation models M1, M2, M3}

\subsubsection{M1}
M1 uses constant innovation covariance followed by scalar dispersion
calibration:
\[
u_t^{(\mathrm{M1})}
=
\gamma_{\mathrm{M1}}L_u z_t,
\]
where \(L_u\) is the Cholesky factor of \(\widehat{\Sigma}_u\), \(z_t\) is a
standardized Student-t shock vector with per-component degrees of freedom,
and \(\gamma_{\mathrm{M1}}\in[0.90,1.50]\) is estimated from training residual
diagnostics.

\subsubsection{M2}
M2 factorizes the innovation into dynamic marginal scale, seasonal cross-mode
dependence, and dispersion calibration:
\begin{equation}
\widetilde{u}_t^{(\mathrm{M2})}
= \operatorname{diag}(\sigma_t^{(\mathrm{EW})})\,L_{q(t)}\,z_t,
\end{equation}
where $\sigma_t^{(\mathrm{EW})}$ is the componentwise EWMA volatility path,
\begin{equation}
(\sigma_{t,k}^{(\mathrm{EW})})^2
= \lambda\,(\sigma_{t-1,k}^{(\mathrm{EW})})^2
  + (1-\lambda)\,\widehat{u}_{t-1,k}^2,
\end{equation}
and $L_{q(t)}$ is the season-specific Cholesky correlation factor. A
coefficient-specific persistence blend (Eq.~\ref{eq:persist_pc}) is then
applied, followed by componentwise dispersion calibration:
$u_t^{(\mathrm{M2})} = \gamma^{(\mathrm{EW})} \odot \breve{u}_t^{(\mathrm{M2})}$.

\subsubsection{M3}
M3 replaces the pure EWMA scale with a GRU-based correction regularized toward
the EWMA backbone. The GRU-to-EWMA ratio is clipped to $[c^{-1},c]$ with
$c=2$, and the final M3 volatility is formed by geometric blending:
\begin{equation}
\sigma_{t,k}^{(\mathrm{M3})}
= \bigl(\sigma_{t,k}^{(\mathrm{EW})}\bigr)^{1-\eta}
  \bigl(\sigma_{t,k}^{(\mathrm{GRU,clip})}\bigr)^\eta,
\end{equation}
where $\eta\in(0,1)$ is a blend weight; in the reported experiments,
$\eta=0.5$. The remainder of the M3 factorization (seasonal correlation, persistence blend,
dispersion calibration) follows M2.

\subsection{Stochastic trajectory branching}

For each ensemble member \(m=1,\ldots,M\), the hybrid samples a branch lead
\(B_m\), interpreted as the last purely dynamical lead. The hybrid path is
\[
\varepsilon_{o+h}^{(\mathrm{hyb},m)}
=
\begin{cases}
\varepsilon_{o+h}^{(\mathrm{dyn},m)}, & h\le B_m,\\[4pt]
\varepsilon_{t_m^*+(h-B_m)}+
\left(1-\dfrac{h-B_m}{\delta}\right)_+
\left(\varepsilon_{o+B_m}^{(\mathrm{dyn},m)}
-\varepsilon_{t_m^*}\right),
& h>B_m,
\end{cases}
\]
before applying the short ramp described below. Here \(t_m^*\) is the selected
donor time, \(\delta\) is the offset-decay length, and
\((x)_+=\max(x,0)\).

The branch lead is sampled as
\[
B_m=h_{\min}+G_m,
\qquad
G_m\sim\mathrm{Geometric}\{1-\exp(-1/\tau)\}
\]
on \(\{0,1,2,\ldots\}\), with truncation at \(H_{\max}\). Thus the fraction of
members still in the dynamical regime at lead \(h\) is approximately
\[
\alpha(h)=
\begin{cases}
1, & h\le h_{\min},\\
\exp\{-(h-h_{\min})/\tau\}, & h>h_{\min},
\end{cases}
\]
up to finite-horizon truncation.

The donor episode is selected by nearest-neighbor matching in the
\(K^*\)-dimensional standardized state among training days whose day-of-year
falls within the CLIM calendar window. Among the \(n_{\mathrm{top}}\) nearest
neighbors, the donor is drawn with softmax probability
\[
\pi_{t^*}
\propto
\exp\!\left(
  -\frac{\|\varepsilon_{o+B_m}^{(\mathrm{dyn},m)}-\varepsilon_{t^*}\|}{T}
\right),
\]
where \(T>0\) is a temperature parameter.

A ramp of \(R\)~days smooths the initial transition. For post-branch offset
\(j=1,\ldots,R\),
\[
\varepsilon_{o+B_m+j}^{(\mathrm{hyb},m)}
=
\beta_j\,\varepsilon_{o+B_m+j}^{(\mathrm{dyn},m)}
+
(1-\beta_j)\,
\varepsilon_{o+B_m+j}^{(\mathrm{donor+offset},m)},
\qquad
\beta_j=1-\frac{j}{R+1}.
\]
For \(j>R\), the hybrid follows the offset-adjusted donor continuation.

\subsubsection{Approximate marginal mixture}
At any fixed lead \(h\), the fraction of members whose branch lead is at
least \(h\) converges, as \(M\to\infty\), to
\begin{equation}
\alpha(h)=
\begin{cases}
1, & h\le h_{\min},\\
\exp\{-(h-h_{\min})/\tau\}, & h>h_{\min}.
\end{cases}
\end{equation}
Outside the \(R\)-day ramp, the fixed-lead predictive distribution therefore
approximates
\begin{equation}
\mathcal P_h^{(\mathrm{Hybrid})}
\approx
\alpha(h)\,\mathcal P_h^{(\mathrm{dyn})}
+ \{1-\alpha(h)\}\,\mathcal P_h^{(\mathrm{CLIM})}.
\end{equation}
During the ramp, the margin also contains convexly interpolated dynamic and
donor states, so exact marginal equivalence does not hold and variation can be
slightly damped. The trajectory construction nevertheless preserves temporal
coherence within each ensemble member.

\section{Additional diagnostics}
\label{app:diagnostics}

This appendix reports additional diagnostics that support and qualify the
main-text results. It addresses three questions: whether the retained EOF state
remains calibrated in coefficient space (not only after grid reconstruction),
whether one-step fitted residuals improve meaningfully after volatility and
dependence standardization, and how sensitive the construction is to region and
regularization level.

\subsection{Coefficient-space calibration}

Table~\ref{tab:pc_pit_cov_summary} examines calibration directly in the
retained EOF coefficients. At short horizons, CLIM attains the lowest PIT
score, as expected from a marginal resampling baseline. The dynamical and
hybrid models remain competitive on mean coefficient-space coverage. At medium
horizons, $\mathrm{M}_3$ and Hybrid-$\mathrm{M}_3$ achieve noticeably lower
failure rates than CLIM and $\mathrm{M}_1$, suggesting that the richer
innovation specification helps stabilize coefficient-space dispersion once the
initial short-range transition has passed.

\begin{table}[t]
\centering
\caption{Coefficient-space PIT and 90\% coverage summaries by horizon group.
PIT score is the Cram\'er-von Mises statistic against a uniform reference
(smaller is better). Mean PC Cov90 is the average empirical 90\% coverage
across retained coefficients. ``Fraction failing'' is the proportion of
coefficients whose coverage falls outside a prespecified tolerance band.}
\label{tab:pc_pit_cov_summary}
\small
\setlength{\tabcolsep}{5pt}
\begin{tabular}{llccc}
\toprule
Horizon & Model & PIT score & Mean PC Cov90 & Fraction failing \\
\midrule
Short & CLIM                       & 0.047 & 0.890 & 0.351 \\
Short & $\mathrm{M}_1$             & 0.070 & 0.829 & 0.557 \\
Short & $\mathrm{M}_3$             & 0.059 & 0.891 & 0.339 \\
Short & Hybrid-$\mathrm{M}_3$      & 0.058 & 0.893 & 0.333 \\
\midrule
Mid   & CLIM                       & 0.053 & 0.902 & 0.267 \\
Mid   & $\mathrm{M}_1$             & 0.068 & 0.870 & 0.402 \\
Mid   & $\mathrm{M}_3$             & 0.058 & 0.927 & 0.175 \\
Mid   & Hybrid-$\mathrm{M}_3$      & 0.057 & 0.926 & 0.171 \\
\bottomrule
\end{tabular}
\end{table}

\subsection{Innovation diagnostics}

Table~\ref{tab:innovation_diag} reports one-step fitted innovation diagnostics
averaged across regions and folds. Among the three dynamical models,
$\mathrm{M}_2$ is closest to nominal on the Mahalanobis summary and has the
highest innovation-level coverage, consistent with its conservative EWMA
dispersion. $\mathrm{M}_3$ shows the highest excess kurtosis, consistent with
a more flexible but occasionally sharper scale path. The seasonal whitening
step removes 82-87\% of off-diagonal cross-mode dependence in all three
models. These one-step results are informative but do not by themselves
determine multi-step field realism; that distinction is important when
interpreting the main-text comparison between $\mathrm{M}_2$ and the other
models.

\begin{table}[t]
\centering
\caption{One-step innovation diagnostics and cross-mode whitening, averaged
across regions and folds. NLL: negative log-likelihood; Mahal: mean
Mahalanobis distance (nominal approximately \(K^*=4\)); mean absolute off-diagonal correlation: mean
off-diagonal correlation before/after seasonal whitening; ARCH rej: ARCH(1)
rejection rate on raw ($z$) and whitened ($z_w$) residuals.}
\label{tab:innovation_diag}
\label{tab:s3_whitening}
\small
\setlength{\tabcolsep}{3.5pt}
\begin{tabular}{lrrrrrrrrr}
\toprule
Model & NLL & Mahal & Cov$_{90}$ & Resid std & Ex.\ kurt
  & $|\bar{r}_{ij}|_{\mathrm{pre}}$ & $|\bar{r}_{ij}|_{\mathrm{post}}$
  & ARCH($z$) & ARCH($z_w$) \\
\midrule
$\mathrm{M}_1$ & 6.91 & 4.40 & 0.857 & 1.048 & 0.56
  & 0.191 & 0.033 & 0.94 & 0.95 \\
$\mathrm{M}_2$ & 6.92 & 3.52 & 0.916 & 0.937 & 0.72
  & 0.192 & 0.025 & 0.60 & 0.69 \\
$\mathrm{M}_3$ & 7.02 & 4.77 & 0.833 & 1.091 & 0.98
  & 0.185 & 0.027 & 0.96 & 0.97 \\
\bottomrule
\end{tabular}
\end{table}

\subsection{Retained-coefficient properties}

Table~\ref{tab:m3_tdf} reports the fitted Student-t degrees of freedom for
innovation shocks by region and retained coefficient. Gulf--South Central shows the
heaviest tails across all four retained modes, consistent with the main-text
finding that this region is the most challenging for calibration. Lag orders
are highly stable across folds: the selected $p$ matches across all four folds
in every region except Central Rockies, where one fold selects $p=3$ instead of
$p=4$.

\begin{table}[t]
\centering
\caption{Fitted Student-t degrees of freedom for innovation shocks by region
and retained coefficient. Lower values indicate heavier tails.}
\label{tab:m3_tdf}
\small
\begin{tabular}{lrrrr}
\toprule
Region & PC 1 & PC\,2 & PC\,3 & PC\,4 \\
\midrule
Upper Midwest & 14.6 & 25.8 & 13.5 & 17.0 \\
Central Rockies & 17.9 & 20.1 & 19.3 & 14.3 \\
Gulf--South Central &  7.6 & 10.6 & 11.8 & 12.5 \\
Southern California & 12.1 & 23.2 & 15.1 & 17.0 \\
Northeast Corridor & 14.7 & 33.2 & 13.1 & 20.0 \\
\bottomrule
\end{tabular}
\end{table}

\subsection{Coverage and calibration details}

Tables~\ref{tab:m4_cov_region}--\ref{tab:s7_warm95} provide detailed
calibration diagnostics from the rolling-origin simulations.
Table~\ref{tab:m4_cov_region} gives regional 90\% coverage deviations from
nominal, Table~\ref{tab:s4_cov_full} reports coverage by horizon for all seven
models, Table~\ref{tab:s6_stability} documents cross-fold coverage stability,
and Table~\ref{tab:s8_seasonal_cov} summarizes seasonal coverage at 30 days.
Table~\ref{tab:s7_warm95} reports W95 area-fraction CRPS by horizon.

\begin{table}[t]
\centering
\caption{Empirical 90\% coverage deviation from nominal by region, averaged
across horizons, origins, and four folds. Positive values indicate
over-coverage; values closer to zero are preferable.}
\label{tab:m4_cov_region}
\small
\setlength{\tabcolsep}{3.5pt}
\begin{tabular}{lrrrrrr}
\toprule
Model & Upper & Central & Gulf & Southern & NE & All \\
\midrule
CLIM & $-$1.4\% & $-$2.8\% & $-$5.7\% & $-$3.6\% & $-$1.3\% & $-$2.9\% \\
$\mathrm{M}_1$ & $+$3.4\% & $+$2.3\% & $+$3.0\% & $+$1.0\% & $+$3.7\% & $+$2.7\% \\
$\mathrm{M}_2$ & $+$2.9\% & $+$2.8\% & $+$3.8\% & $+$1.5\% & $+$2.5\% & $+$2.7\% \\
$\mathrm{M}_3$ & $-$0.2\% & $+$1.3\% & $-$1.2\% & $-$1.0\% & $+$2.5\% & $+$0.3\% \\
Hyb-$\mathrm{M}_1$ & $+$1.4\% & $+$0.7\% & $+$0.4\% & $-$0.3\% & $+$2.0\% & $+$0.8\% \\
Hyb-$\mathrm{M}_2$ & $+$1.6\% & $+$1.1\% & $+$1.4\% & $+$0.1\% & $+$1.3\% & $+$1.1\% \\
Hyb-$\mathrm{M}_3$ & $-$1.2\% & $-$0.3\% & $-$2.1\% & $-$1.8\% & $+$1.1\% & $-$0.9\% \\
\bottomrule
\end{tabular}
\end{table}

\begin{table}[t]
\centering
\caption{Empirical 90\% coverage by horizon for all seven models.}
\label{tab:s4_cov_full}
\small
\setlength{\tabcolsep}{3.2pt}
\begin{tabular}{lrrrrrrrr}
\toprule
Model & $h{=}1$ & $h{=}3$ & $h{=}7$ & $h{=}14$ & $h{=}30$ & $h{=}60$ & $h{=}90$ & $h{=}120$ \\
\midrule
CLIM & 0.881 & 0.875 & 0.867 & 0.860 & 0.876 & 0.870 & 0.874 & 0.862 \\
$\mathrm{M}_1$ & 0.932 & 0.923 & 0.929 & 0.921 & 0.929 & 0.924 & 0.933 & 0.921 \\
$\mathrm{M}_2$ & 0.931 & 0.932 & 0.937 & 0.923 & 0.930 & 0.926 & 0.926 & 0.912 \\
$\mathrm{M}_3$ & 0.909 & 0.902 & 0.902 & 0.896 & 0.902 & 0.903 & 0.911 & 0.899 \\
Hyb-$\mathrm{M}_1$ & 0.932 & 0.923 & 0.929 & 0.916 & 0.917 & 0.899 & 0.884 & 0.866 \\
Hyb-$\mathrm{M}_2$ & 0.931 & 0.932 & 0.937 & 0.918 & 0.921 & 0.900 & 0.891 & 0.860 \\
Hyb-$\mathrm{M}_3$ & 0.909 & 0.902 & 0.902 & 0.892 & 0.896 & 0.887 & 0.881 & 0.861 \\
\bottomrule
\end{tabular}
\end{table}

\begin{table}[t]
\centering
\caption{Cross-fold stability of 90\% coverage (mean +/- std across four
folds). The largest cross-fold standard deviation among the displayed entries
is 0.029 (CLIM, Gulf--South Central).}
\label{tab:s6_stability}
\small
\begin{tabular}{lrrrr}
\toprule
Region & CLIM & $\mathrm{M}_3$ & Hyb-$\mathrm{M}_2$ & Hyb-$\mathrm{M}_3$ \\
\midrule
Upper Midwest & $.887\!\pm\!.017$ & $.898\!\pm\!.011$ & $.916\!\pm\!.004$ & $.889\!\pm\!.010$ \\
Central Rockies & $.872\!\pm\!.009$ & $.913\!\pm\!.009$ & $.911\!\pm\!.010$ & $.897\!\pm\!.013$ \\
Gulf--South Central & $.843\!\pm\!.029$ & $.888\!\pm\!.008$ & $.914\!\pm\!.007$ & $.879\!\pm\!.005$ \\
Southern California & $.864\!\pm\!.017$ & $.890\!\pm\!.018$ & $.901\!\pm\!.014$ & $.882\!\pm\!.019$ \\
Northeast Corridor & $.887\!\pm\!.012$ & $.925\!\pm\!.009$ & $.913\!\pm\!.007$ & $.911\!\pm\!.005$ \\
\bottomrule
\end{tabular}
\end{table}

\begin{table}[t]
\centering
\caption{Seasonal 90\% coverage at h=30 days. Values are grouped by target
season.}
\label{tab:s8_seasonal_cov}
\small
\begin{tabular}{lrrrrr}
\toprule
Model & DJF & MAM & JJA & SON & Range \\
\midrule
CLIM               & 0.881 & 0.909 & 0.825 & 0.890 & 0.084 \\
$\mathrm{M}_1$     & 0.924 & 0.917 & 0.954 & 0.923 & 0.037 \\
$\mathrm{M}_3$     & 0.907 & 0.904 & 0.917 & 0.879 & 0.038 \\
Hyb-$\mathrm{M}_3$ & 0.899 & 0.905 & 0.913 & 0.869 & 0.044 \\
\bottomrule
\end{tabular}
\end{table}

\begin{table}[t]
\centering
\caption{Warm-exceedance area-fraction CRPS by horizon. Lower is better. CLIM
remains the strongest marginal-tail reference at most horizons beyond one day,
while the dynamical models improve W95 at h=1.}
\label{tab:s7_warm95}
\small
\setlength{\tabcolsep}{3.2pt}
\begin{tabular}{lrrrrrrrr}
\toprule
Model & $h{=}1$ & $h{=}3$ & $h{=}7$ & $h{=}14$ & $h{=}30$ & $h{=}60$ & $h{=}90$ & $h{=}120$ \\
\midrule
CLIM               & 0.074 & 0.085 & 0.094 & 0.094 & 0.081 & 0.083 & 0.101 & 0.090 \\
$\mathrm{M}_1$     & 0.054 & 0.092 & 0.110 & 0.114 & 0.105 & 0.108 & 0.124 & 0.108 \\
$\mathrm{M}_3$     & 0.051 & 0.087 & 0.103 & 0.107 & 0.098 & 0.102 & 0.117 & 0.102 \\
Hyb-$\mathrm{M}_3$ & 0.051 & 0.087 & 0.103 & 0.106 & 0.096 & 0.093 & 0.109 & 0.090 \\
\bottomrule
\end{tabular}
\end{table}

\begin{table}[t]
\centering
\caption{Per-coefficient member-level autocorrelation at h=1 and h=7 days
for selected models. Dynamic models preserve PC 1 autocorrelation very closely
at h=1; higher-order PCs show faster degradation.}
\label{tab:m7_perpc_acf}
\small
\setlength{\tabcolsep}{3.5pt}
\begin{tabular}{lrrrrrrrr}
\toprule
 & \multicolumn{4}{c}{h=1} & \multicolumn{4}{c}{h=7} \\
\cmidrule(lr){2-5} \cmidrule(lr){6-9}
Model & PC1 & PC2 & PC3 & PC4 & PC1 & PC2 & PC3 & PC4 \\
\midrule
Observed        & 0.961 & 0.577 & 0.508 & 0.527 & 0.863 & 0.236 & 0.149 & 0.202 \\
\midrule
CLIM            & 0.850 & 0.171 & 0.083 & 0.180 & 0.846 & 0.172 & 0.078 & 0.181 \\
$\mathrm{M}_1$ & 0.954 & 0.579 & 0.482 & 0.475 & 0.832 & 0.172 & 0.073 & 0.158 \\
$\mathrm{M}_3$ & 0.959 & 0.579 & 0.461 & 0.462 & 0.846 & 0.184 & 0.077 & 0.160 \\
Hyb-$\mathrm{M}_3$ & 0.959 & 0.579 & 0.461 & 0.462 & 0.846 & 0.184 & 0.077 & 0.160 \\
\bottomrule
\end{tabular}
\end{table}

\begin{table}[t]
\centering
\caption{Mean absolute member-level autocorrelation error by horizon. Dynamic models are
substantially closer to observed at the one-day horizon; all models converge by
h>=60.}
\label{tab:s10_acf_error}
\small
\begin{tabular}{lrrrrrrr}
\toprule
Model & $h{=}1$ & $h{=}3$ & $h{=}7$ & $h{=}14$ & $h{=}30$ & $h{=}60$ & $h{=}90$ \\
\midrule
CLIM            & 0.323 & 0.133 & 0.121 & 0.119 & 0.106 & 0.103 & 0.099 \\
$\mathrm{M}_1$ & 0.071 & 0.123 & 0.128 & 0.124 & 0.116 & 0.115 & 0.098 \\
$\mathrm{M}_3$ & 0.071 & 0.122 & 0.122 & 0.121 & 0.111 & 0.112 & 0.099 \\
Hyb-$\mathrm{M}_3$ & 0.071 & 0.122 & 0.122 & 0.120 & 0.110 & 0.111 & 0.098 \\
\bottomrule
\end{tabular}
\end{table}

\subsection{Autocorrelation and hyperparameter diagnostics}

Tables~\ref{tab:m7_perpc_acf} and~\ref{tab:s10_acf_error} report
autocorrelation diagnostics in the retained EOF coefficients. At h=1, the
dynamical models reproduce PC 1 autocorrelation very closely, whereas CLIM
preserves substantially less dependence beyond the leading mode. By h>=60,
the differences become much smaller, consistent with the main-text conclusion
that the dynamical advantage is primarily a short- to medium-range phenomenon. Table~\ref{tab:m9_hybrid_tau} gives the seasonal breakdown of $\tau$ for
$\mathrm{M}_3$, and Table~\ref{tab:s1_tuning} records the tuned EWMA
smoothing parameters and GRU lookback lengths across folds. The main tuning
parameters are reasonably stable across folds, especially the EWMA smoothing
parameter $\lambda$.

\begin{table}[t]
\centering
\caption{Fitted hybrid regularization time scale \(\tau\) (days) for
\(\mathrm{M}_3\) by region, with seasonal breakdown. Larger \(\tau\) implies
slower decay from the parametric low-rank generator toward state-matched
climatological continuations; it should not be read as a direct physical
predictability horizon. Values are averaged over rolling folds and day-of-year
bins.}
\label{tab:m9_hybrid_tau}
\small
\begin{tabular}{lrrrrr}
\toprule
Region & Min \(\tau\) & Max \(\tau\) & Mean \(\tau\) & Winter \(\tau\) & Summer \(\tau\) \\
\midrule
Upper Midwest & 30 & 720 & 217 & 166 & 176 \\
Central Rockies & 30 & 720 & 173 & 198 & 160 \\
Gulf--South Central & 30 & 720 & 262 & 290 & 174 \\
Southern California & 30 & 720 & 310 & 258 & 335 \\
Northeast Corridor & 30 & 720 & 180 & 180 & 175 \\
\bottomrule
\end{tabular}
\end{table}

\begin{table}[t]
\centering
\caption{Tuned EWMA lambda and GRU lookback L across the four
rolling-origin folds.}
\label{tab:s1_tuning}
\small
\begin{tabular}{lrr}
\toprule
Region & $\lambda$ across folds & GRU $L$ across folds \\
\midrule
Upper Midwest & 0.95, 0.95, 0.95, 0.95 & 45, 45, 45, 15 \\
Central Rockies & 0.95, 0.95, 0.95, 0.98 & 15, 15, 15, 15 \\
Gulf--South Central & 0.93, 0.93, 0.95, 0.95 & 15, 15, 30, 30 \\
Southern California & 0.97, 0.97, 0.95, 0.99 & 15, 30, 30, 45 \\
Northeast Corridor & 0.98, 0.99, 0.97, 0.99 & 45, 30, 30, 15 \\
\bottomrule
\end{tabular}
\end{table}


\section{Experimental design and implementation details}
\label{app:implementation}

This section records the experimental protocol, model settings, and evaluation
configuration used in the main text. All estimated quantities are fitted within
each training fold and then applied unchanged out of sample.

\subsection{Core experimental settings}

Table~\ref{tab:core_settings} collects the main settings in one place.

\begin{table}[t]
\centering
\caption{Core experimental settings used in the reported results.}
\label{tab:core_settings}
\small
\setlength{\tabcolsep}{5pt}
\begin{tabularx}{\textwidth}{lX}
\toprule
Component & Setting \\
\midrule
Domains & Upper Midwest, Southern California, Gulf--South Central, Central Rockies, Northeast Corridor \\
Grid size & $64\times64$ per region ($D=4096$) \\
Training/test design & expanding-window rolling origin, 10-year test blocks, 5-year step \\
Folds & 4 folds: 2001-2010, 2006-2015, 2011-2020, 2016-2025 \\
EOF stability blocks & 20-year rolling sub-blocks, 10-year step \\
Deterministic mean/scale & 6 harmonic pairs, A/B/C/D model selection, optional Theil--Sen drift \\
Mean/scale regularization & ridge penalty $10^{-3}$, minimum scale floor $5\times10^{-3}$ \\
Dependence model & seasonal ridge-regularized VARX, lag grid $\{1,\dots,10\}$, BIC selection \\
VARX regularization & ridge penalty \(2\times10^{-2}\), target spectral radius \(<0.97\); inverse-variance ridge target (strength 3.0); mode \(k\) uses \(p_k=\max\{p-k+1,1\}\) \\
EWMA \(\lambda\) tuning & grid \(\{0.90,0.93,0.95,0.97,0.98,0.99\}\), inner train/validation split \\
GRU volatility & lookback tuned on \(\{15,30,45\}\), hidden size on \(\{64,128\}\), 200 epochs, learning rate \(3\times10^{-4}\), weight decay \(5\times10^{-4}\) \\
GRU regularization & ratio clip \(c=2\), blend weight 0.5 relative to EWMA baseline \\
Shock distribution & Student-\(t\), per-component df from whitened residuals \\
Persistence blend & per-coefficient \(\boldsymbol{\rho}=(0.1,0,0,0)\) for \(\mathrm{M}_2\) and \(\mathrm{M}_3\) \\
Dispersion calibration & M1 scalar \(\gamma\in[0.90,1.50]\); M2 componentwise \(\gamma_k\in[0.90,1.60]\); \(\mathrm{M}_3\) componentwise calibration after EWMA-regularized GRU scale estimation \\
Short-horizon inflation & \(\mathrm{M}_3\) only; innovation-scale multipliers 1.10, 1.05, and 1.02 at leads 1, 2, and 3 \\
Simulation horizons & \(\{1,3,7,14,30,60,90,120\}\) days \\
Ensemble size & \(M=100\) \\
Origin-spacing settings & nominal stride 90 days with 7-day jitter; valid origins require \(H_{\max}=120\) future days \\
Dependence-aware scoring & pixel subset size 512 \\
Coverage evaluation & levels $(0.50,0.80,0.90,0.95,0.99)$ \\
Coefficient-space coverage & levels $(0.80,0.90,0.95,0.99)$ \\
Hybrid tuning grid & $\tau\in\{30,60,\ldots,360,540,720\}$ days, 24 DOY bins \\
Trajectory branching & $h_{\min}=7$, ramp $R=5$, offset decay $\delta=18$, $n_{\mathrm{top}}=3$, temperature $T=1.5$ \\
\bottomrule
\end{tabularx}
\end{table}

The constants in Table~\ref{tab:core_settings} define the training-only
regularization and simulation protocol used for all regions and folds. They
were held fixed for out-of-sample scoring and were not re-tuned separately by
region after inspecting test scores. The most important free hybrid parameter,
\(\tau\), is fitted from training data by day-of-year bin; the remaining
branching constants control splice smoothness and state-matched donor
selection. These constants are development-stage design choices rather than
estimated scientific parameters; their exact command-line values are preserved
in the fixed run configuration. Sensitivity analysis over these secondary
constants is a useful extension, but the main comparisons in the paper use one
fixed protocol.

\subsection{Rolling-origin design}
\label{app:cv}

We use an expanding-window rolling-origin design with calendar-year test
blocks. For a fold with test-start year $y$, the training set contains all
observations with year $\le y-1$, and the test set contains all observations
from year $y$ through year $y+9$. In the reported experiments, the first test
year is 2001, the test-block length is 10 years, the step between test origins
is 5 years, and the minimum training size is 7300 daily observations. This
yields four folds per region (Table~\ref{tab:cv_folds}). Because the test
periods overlap, cross-fold variability should be interpreted as a robustness
summary rather than as independent-sample variation.

\begin{table}[t]
\centering
\caption{Rolling-origin folds used in the experiments.}
\label{tab:cv_folds}
\small
\begin{tabular}{ccc}
\toprule
Fold & Training period & Test period \\
\midrule
1 & 1940-2000 & 2001-2010 \\
2 & 1940-2005 & 2006-2015 \\
3 & 1940-2010 & 2011-2020 \\
4 & 1940-2015 & 2016-2025 \\
\bottomrule
\end{tabular}
\end{table}

\subsection{Native coverage and filling of external covariates}
\label{app:covariate_coverage}

Table~\ref{tab:covariate_coverage} records when each monthly covariate is
natively observed and which parts of the 1940--2025 analysis interval are
filled by the preprocessing convention described in Sect.~\ref{sec:covariates}.
The table is part of the data definition: the filled portions are retained only
as descriptive controls and are not treated as observed historical forcing.

\begin{table}[t]
\centering
\caption{Native monthly coverage and filled intervals for the external
covariates in the 1940--2025 study record. ``Leading fill'' denotes back-filling
with the first available monthly value; ``trailing fill'' denotes
forward-filling with the last available monthly value. Dates refer to the documented source-series coverage used to construct the monthly covariate table.}
\label{tab:covariate_coverage}
\small
\setlength{\tabcolsep}{6pt}
\begin{tabularx}{\textwidth}{lXXX}
\toprule
Covariate & Native source coverage used & Leading fill in study period & Trailing fill in study period \\
\midrule
CO$_2$ & March 1958--December 2025 & January 1940--February 1958 & none \\
ONI & January 1950--December 2025 & January 1940--December 1949 & none \\
AMO & January 1856--January 2023 & none & February 2023--December 2025 \\
\bottomrule
\end{tabularx}
\end{table}

\subsection{Deterministic coefficient models}
\label{app:mean_scale_hypers}

For each retained EOF coefficient, the deterministic mean is selected from a
candidate family of increasing complexity: intercept only; intercept plus
harmonics; harmonics plus covariates; and harmonics plus covariates plus
Theil--Sen drift \citep{theil1950rank,sen1968estimates}. Model selection uses a
training-only time-series
cross-validation criterion. The scale model is log-linear in the same harmonic
and covariate features. Ridge penalties of $10^{-3}$ are applied to both mean
and scale, with a minimum scale floor of $5\times10^{-3}$.

In the reported experiments, the external covariates are aligned with a fixed
causal lag of \(\ell=1\) day and are standardized using training data only.
The monthly covariate table is first aggregated to month-start timestamps,
forward-filled across missing months, and back-filled once for any leading
missing values. It is then aligned to daily ERA5 dates by backward as-of
matching and shifted by one day. Under this construction, each simulation date
uses the most recently available one-day-lagged monthly value; it is not a
strict previous-calendar-month assignment for every day.

\subsection{Dependence model}
\label{app:varx_hypers}

The standardized coefficient vector is propagated with a seasonal
ridge-regularized VARX model: separate intercept and autoregressive coefficient
matrices are estimated for each meteorological season (DJF, MAM, JJA, SON).
Lag selection is performed over the grid \(\{1,2,\dots,10\}\) using BIC. The
ridge penalty is \(2\times10^{-2}\), exogenous inputs enter in lagged form, and
the target maximum spectral radius is 0.97. The inverse-variance ridge target
has strength 3.0. Under the target-specific descending lag cap, retained mode \(k\) uses
\(p_k=\max\{p-k+1,1\}\).

\subsection{Innovation specification}
\label{app:innovation_hypers}

For $\mathrm{M}_2$ and $\mathrm{M}_3$, simulated innovations are factorized as
\[
\widetilde{u}_t
=
\gamma\odot\bigl[\mathrm{diag}(\sigma_t)\,L_{\mathrm{season}(t)}\,z_t\bigr],
\]
where \(\sigma_t\) is the componentwise volatility path,
\(L_{\mathrm{season}(t)}\) is the season-specific Cholesky correlation factor,
\(\gamma\in\mathbb{R}^{K^*}_{>0}\) is the dispersion-calibration vector, and
\(z_t\) is a Student-t shock vector with per-component degrees of freedom.

For $\mathrm{M}_2$, the EWMA smoothing parameter \(\lambda\) is tuned on an
inner training/validation split over the grid
\[
\{0.90,0.93,0.95,0.97,0.98,0.99\}.
\]
The M2 EWMA scale is capped at three times its reference scale before final
calibration.

For $\mathrm{M}_3$, the GRU is trained to predict a log-scale correction to
the concurrent EWMA baseline. The input at each training case is a lookback
window of standardized residuals divided by the EWMA scale, augmented with
seasonal Fourier features. The target is the clipped log correction
\[
\frac{1}{2}
\left[
\log\{\widehat{u}_{t,k}^2+\epsilon\}
-
\log\{(\sigma_{t,k}^{(\mathrm{EW})})^2+\epsilon\}
\right],
\]
and the GRU is fitted by mean squared error over retained coefficients. The
lookback length is tuned on \(\{15,30,45\}\), the hidden dimension on
\(\{64,128\}\), and the final GRU-to-EWMA ratio is constrained to
\([c^{-1},c]\) with \(c=2\). The clipped GRU estimate is geometrically averaged
with the EWMA scale using blend weight 0.5. A small M3-only short-horizon
innovation-scale inflation is applied at leads 1--3 with multipliers 1.10,
1.05, and 1.02.

A coefficient-specific innovation-persistence blend with
\(\boldsymbol{\rho}=(0.1,0,0,0)\) is applied to \(\mathrm{M}_2\) and
M3 before final dispersion calibration, providing mild
persistence stabilization for the dominant mode only. For \(\mathrm{M}_1\),
innovations are drawn from
\[
u_t^{(\mathrm{M1})}
=
\gamma_{\mathrm{M1}} L_u z_t,
\]
where \(L_u\) is the Cholesky factor of the full training innovation covariance
\(\widehat{\Sigma}_u\), \(z_t\) is a Student-t shock vector, and
\(\gamma_{\mathrm{M1}}\in[0.90,1.50]\) is a scalar calibration factor.

\subsection{Climatology and hybrid models}
\label{app:clim_hybrid}

The CLIM baseline resamples contiguous training episodes matched by day of year
using a $\pm 15$-day window, expanded in steps of 5 days up to $\pm 60$ when
the candidate pool falls below 250 start dates. Because episodes are sampled as
contiguous blocks, the benchmark preserves training-set temporal dependence
rather than treating days independently.

Hybrid simulations combine a dynamical model with CLIM using trajectory-level
branching. The branch distribution is parameterized by a fitted time scale
\(\tau\), with dynamical-member fraction approximately
\[
\alpha(h)=
\begin{cases}
1, & h\le h_{\min},\\
\exp\{-(h-h_{\min})/\tau\}, & h>h_{\min}.
\end{cases}
\]
The parameter \(\tau\) is estimated from training data over the grid
\(\{30,60,\ldots,360,540,720\}\) days, using 24 day-of-year bins with
smoothness regularization and fitting restricted to horizons \(h\ge 7\).

\subsection{Monte Carlo evaluation}
\label{app:mc_eval}

Ensemble simulations are generated in EOF coefficient space and reconstructed
to the grid only as needed for evaluation. The simulation horizons are
\(\{1,3,7,14,30,60,90,120\}\) days and the ensemble size is \(M=100\).
Evaluation origins are selected using a nominal 90-day stride with 7-day
jitter. Valid origins are required to have at least \(H_{\max}=120\) future
days available, so the 90-day setting controls origin spacing rather than
limiting the simulated trajectory length. Dependence-aware field scores are
computed on a fixed fold-specific subset of 512 grid cells.

\subsection{Implementation notes}
\label{app:implementation_notes}

PCA, deterministic fitting, and simulation are all carried out primarily in EOF
coefficient space. Reconstruction to the $64\times64$ grid is performed only
for verification and summary output. This avoids materializing full grid
trajectories during simulation and substantially reduces memory use. When GPU
hardware is available, GRU volatility training and multi-horizon Monte Carlo
simulation are accelerated accordingly; the remaining components are fitted on
CPU.

\subsubsection{Computational cost}
For a single region and fold, the full pipeline completes in approximately
15-25 minutes on a workstation with a single NVIDIA A100 GPU and 32-core CPU.
GRU volatility training accounts for roughly 40\% of the wall time; the
remaining components each complete in under two minutes. Because the model
operates entirely in $K^*=4$ dimensional coefficient space rather than the
native $D=4096$ grid, peak memory usage is below 8\,GB even for the longest
training folds. The full experiment (five regions $\times$ four folds $\times$
seven models) runs overnight on a single multi-GPU node. These costs are
substantially lower than those of deep-learning weather emulators and
comparable to multisite weather generators operating at similar spatial
resolution.


\clearpage
\section*{Code availability}
The analysis code, preprocessing scripts, fixed run
configuration, random seeds, and pinned software environment will be deposited
in a versioned FAIR-aligned public repository before publication. The final
repository DOI will be added to the manuscript and reference list when the
archive is released. During peer review, the versioned code archive can be
provided to the editor and referees on request.

\section*{Data availability}
ERA5 reanalysis data are available from the Copernicus
Climate Data Store and are cited in the reference list. The monthly CO$_2$, ONI,
and AMO indices used in this study are available from NOAA sources and are cited
in the reference list. No original observational data were generated by this
study.

\section*{Author contributions}
VK developed the methodology and software, performed the
experiments, analyzed the results, and wrote the initial manuscript draft. MLS
provided supervision, conceptual guidance, methodological feedback, and
manuscript review and editing.

\section*{Competing interests}
The authors declare that they have no conflict of interest.

\section*{Acknowledgements}
This material was based upon work supported by the U.S. Department of Energy,
Office of Science, Office of Advanced Scientific Computing Research (ASCR) under
Contract DE-AC02-06CH11357. Generative AI tools were used for language editing
and bibliographic consistency checks; no scientific analyses or numerical
results were generated by these tools.

\section*{Financial support}
This work was supported by the U.S. Department of Energy,
Office of Science, Office of Advanced Scientific Computing Research (ASCR) under
Contract DE-AC02-06CH11357.

\end{document}